\documentclass[fleqn,usenatbib]{mnras}

\usepackage{newtxtext,newtxmath}

\usepackage[T1]{fontenc}

\DeclareRobustCommand{\VAN}[3]{#2}
\let\VANthebibliography\thebibliography
\def\thebibliography{\DeclareRobustCommand{\VAN}[3]{##3}\VANthebibliography}

\usepackage{graphicx}	
\usepackage{amsmath}	
\usepackage{cleveref}	
\crefname{figure}{Fig.}{Figs.}
\crefname{equation}{equation}{equations}

\usepackage{adjustbox}

\newcommand{\um}{\mu\mathrm{m}}
\newcommand{\id}{GNWY-7379420231}

\newcommand{\orcidsymb}[2]{#1\href{http://orcid.org/#2}{\adjustbox{trim={-.15\width} {0\height} {-.15\width} {0\height},clip}{\includegraphics[height=10pt]{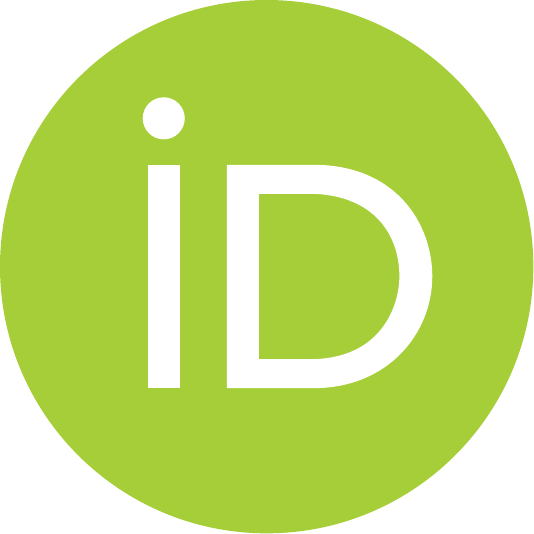}}}}

\title[The $2175${\AA} UV Bump in the EoR]{Detection of the $2175${\AA} UV Bump at $z>7$: Evidence for Rapid Dust Evolution in a Merging Reionization-Era Galaxy}

\author[K. Ormerod et al.]{\orcidsymb{Katherine Ormerod}{0000-0003-2000-3420}$^{1}$\thanks{E-mail: \href{mailto:arikorme@ljmu.ac.uk}{arikorme@ljmu.ac.uk}},
\orcidsymb{Joris Witstok}{0000-0002-7595-121X}$^{2,3}$,
\orcidsymb{Renske Smit}{0000-0001-8034-7802}$^{1}$,
\orcidsymb{Anna de Graaff}{0000-0002-2380-9801}${^4}$,
\orcidsymb{Jakob M. Helton}{0000-0003-4337-6211}$^{5}$,\newauthor
\orcidsymb{Michael V.\ Maseda}{0000-0003-0695-4414}${^6}$,
\orcidsymb{Irene Shivaei}{0000-0003-4702-7561}${^7}$,
\orcidsymb{Andrew J.\ Bunker }{0000-0002-8651-9879}${^8}$,
\orcidsymb{Stefano Carniani}{0000-0002-6719-380X}${^9}$,
\newauthor
\orcidsymb{Francesco D'Eugenio}{0000-0003-2388-8172}$^{10,11}$,
\orcidsymb{Rachana Bhatawdekar}{0000-0003-0883-2226}${^{13}}$,
\orcidsymb{Jacopo Chevallard}{0000-0002-7636-0534}${^{8}}$,
\orcidsymb{Marijn Franx}{0000-0002-8871-3026}${^{14}}$,
\newauthor
\orcidsymb{Nimisha Kumari}{0000-0002-5320-2568}${^{15}}$,
\orcidsymb{Roberto Maiolino}{0000-0002-4985-3819}$^{10,11, 12}$,
\orcidsymb{Pierluigi Rinaldi}{0000-0002-5104}${^{5}}$,
\orcidsymb{Brant Robertson}{0000-0002-4271-0364}${^{16}}$,
\newauthor
\orcidsymb{Sandro Tacchella}{0000-0002-8224-4505}$^{10,11}$
\\
$^{1}$Astrophysics Research Institute, Liverpool John Moores University, 146 Brownlow Hill, Liverpool, L3 5RF, UK \\
$^{2}$Cosmic Dawn Center (DAWN), Copenhagen, Denmark \\
${^3}$Niels Bohr Institute, University of Copenhagen, Jagtvej 128, DK-2200, Copenhagen, Denmark\\
${^4}$Max-Planck-Institut f\"ur Astronomie, K\"onigstuhl 17, D-69117, Heidelberg, Germany\\
${^5}$Steward Observatory, University of Arizona, 933 N. Cherry Avenue, Tucson, AZ 85721, USA\\
$^6$Department of Astronomy, University of Wisconsin-Madison, 475 N. Charter St., Madison, WI 53706, USA\\
${^7}$Centro de Astrobiología (CAB), CSIC-INTA, Ctra. de Ajalvir km 4, Torrejón de Ardoz, E-28850, Madrid, Spain\\
${^8}$Department of Physics, University of Oxford, Denys Wilkinson Building, Keble Road, Oxford OX1 3RH, UK\\
${^9}$Scuola Normale Superiore, Piazza dei Cavalieri 7, I-56126 Pisa, Italy\\
$^{10}$Kavli Institute for Cosmology, University of Cambridge, Madingley Road, Cambridge, CB3 0HA, UK\\
$^{11}$Cavendish Laboratory, University of Cambridge, 19 JJ Thomson Avenue, Cambridge, CB3 0HE, UK\\
${^{12}}$Department of Physics and Astronomy, University College London, Gower Street, London WC1E 6BT, UK\\
${^{13}}$European Space Agency (ESA), European Space Astronomy Centre (ESAC), Camino Bajo del Castillo s/n, 28692 Villanueva de la Cañada, Madrid, Spain\\
${^{14}}$Leiden Observatory, Leiden University, NL-2300 RA Leiden, Netherlands\\
${^{15}}$AURA for European Space Agency, Space Telescope Science Institute, 3700 San Martin Drive. Baltimore, MD, 21210\\
${^{16}}$Department of Astronomy and Astrophysics University of California, Santa Cruz, 1156 High Street, Santa Cruz CA 96054, USA\\
}

\date{Accepted XXX. Received YYY; in original form ZZZ}

\pubyear{\the\year{}}

\begin{document}
\label{firstpage}
\pagerange{\pageref{firstpage}--\pageref{lastpage}}
\maketitle

\begin{abstract}
Dust is a fundamental component of the interstellar medium (ISM) within galaxies, as dust grains are highly efficient absorbers of UV and optical photons. Accurately quantifying this obscuration is crucial for interpreting galaxy spectral energy distributions (SEDs). The extinction curves in the Milky Way (MW) and Large Magellanic Cloud (LMC) exhibit a strong feature known as the $2175$\AA\ UV bump, most often attributed to small carbonaceous dust grains. This feature was recently detected in faint galaxies out to $z=7.55$, suggesting rapid formation channels.
Here we report the detection of a strong UV bump in a luminous Lyman-break galaxy at $z_\mathrm{prism}=7.11235 $, GNWY-7379420231,   through observations taken as part of the NIRSpec Wide GTO survey. We fit a dust attenuation curve that is consistent with the MW extinction curve within $1\sigma$, in a galaxy just $\sim 700$ Myr after the Big Bang. 
From the integrated spectrum, we infer a young mass-weighted age ($t_\star \sim 22-59$ Myr) for this galaxy, however spatially resolved SED fitting unveils the presence of an older stellar population ($t_\star \sim 252$ Myr). Furthermore, morphological analysis provides evidence for a potential merger. The underlying older stellar population suggests the merging system could be pre-enriched, with the dust illuminated by a merger-induced starburst. Moreover, turbulence driven by stellar feedback in this bursty region may be driving PAH formation through top-down shattering. 
The presence of a UV bump in GNWY-7379420231 solidifies growing evidence for the rapid evolution of dust properties within the first billion years of cosmic time. 
\end{abstract}




\section{Introduction}

Dust is a fundamental component of the interstellar medium (ISM) within galaxies, and affects the observed spectral energy distribution (SED) of galaxies over a broad wavelength range. Dust grains absorb approximately half of the optical and ultraviolet (UV) light and re-emit the absorbed energy as infrared light, which has important implications on the observational properties of galaxies \citep{kennicutt_evans_2012, schneider_formation_2023}.  
The dust attenuation curve of a galaxy describes how the integrated luminosity of a galaxy is affected by absorption and scattering of photons along the line of sight (LOS) due to dust in the ISM, and results from a combination of dust grain properties, dust content, and the spatial distribution of dust \citep[e.g.,][]{Calzetti_13, Salim_2020, Markov23, Markov24}. An understanding of a galaxy's attenuation curve is crucial for the derivation of robust physical parameters, which can vary significantly depending on the assumed attenuation curve \citep[e.g.,][]{kriek_conroy_2013, Shivaei_reddy_2020, Reddy_2015, Salim_2016, Salim_2020, Markov23}. 

Dust in galaxies can be characterized using both extinction curves, measured along sightlines to individual stars, and attenuation curves describing the integrated light of galaxies. Attenuation curves incorporate effects arising from the star-dust geometry within galaxies, such as scattering back into the line of sight, and contribution from unobscured stars \citep{Narayanan_2018, Salim_2020}. 
Common examples include the Calzetti attenuation curve \citep{Calzetti_1994, Calzetti00} derived for local starburst galaxies, and the Milky Way (MW) \citep{Cardelli_1989}, Small Magellanic Cloud (SMC), and the Large Magellanic Cloud (LMC) extinction curves \citep{Fitzpatrick_1986, Gordon_2003, Gordon_2024}. The MW extinction curve exhibits a `UV bump' feature at $2175$\AA, while the LMC extinction curve contains a weaker UV bump, but stronger far ultraviolet (FUV) rise than the MW curve. In general, these dust curves vary in the slope in the UV/optical range, and the presence (or absence) of a UV bump. It has also been shown that galaxies in the local universe exhibit a wide range of dust attenuation curves, which can, for instance, be parameterized with the Salim dust curve \citep{Salim2018}. This parameterization is a modified Calzetti curve which allows the slope of the curve to vary, and allows for the presence or absence of a UV bump. 
Alternative dust attenuation curves include the flexible, two-component \citet{CF00} model, and the \citet{Noll_2009} model which modifies the Calzetti curve by including a Lorentzian-like UV bump. 
The UV bump strength is thought to vary with the slope of the dust attenuation curve, such that flatter attenuation curves display a weaker bump strength \citep{kriek_conroy_2013, Narayanan_2018}. 
It must be noted that this correlation is not seen in all studies. For example, \citet{Buat2011, Buat2012} find no clear relationship between the slope of the attenuation curve and UV bump strength.
It has also been suggested that for galaxies with fixed optical depth, galaxies with higher metallicities have flatter attenuation curves but stronger UV bump strengths \citep{Shivaei_reddy_2020}, indicating a lower prevalence of the dust grains responsible for the UV bump at low metallicity. However, this relationship is not observed universally, with studies such as \citet{Salim_2020} suggesting that star-dust geometry and radiative transfer effects can also affect the UV bump strength.

The UV bump is a strong feature seen in the dust attenuation curve of some galaxies, and was first detected in MW sightlines by \citet{stecher_1965}.
The origin of this feature is not well known \citep{Draine_1989}, and was initially suggested to be caused by graphite \citep{Stecher_1965b}. Today, this feature is most commonly attributed to nanoparticles containing aromatic carbon (C), such as polycyclic aromatic hydrocarbons (PAHs) \citep[e.g.,][]{Joblin_1992, Bradley_2005, Shivaei_2022}, or nano-sized graphite grains \citep{li_draine2001}. Other interpretations, including a random arrangement of microscopic sp$^2$ carbon chips, have also been proposed \citep{Popoular_2009}.
PAHs are hydrocarbon molecules with C atoms arranged in a honeycomb structure of fused aromatic rings with peripheral H atoms, and are abundant in the ISM \citep{Tielens_2008}. 

Beyond the local Universe, this feature has only been seen spectroscopically in metal-enriched galaxies at $0.01 \leq z \leq 3$ \citep[e.g.,][]{Noll_2007, Noll_2009, Shivaei_2022}, and was first seen in a galaxy at $z > 3$ in the spectrum of  JADES-GS-z6-0 at $z = 6.71$ \citep{Witstok_2023_Bump}, with tentative evidence of a higher peak wavelength than that typically observed within the MW, suggesting a differing mixture of carbonaceous grains \citep{blasberger2017}. The UV bump has since been detected in individual galaxies at redshifts up to $z\sim7.55$ \citep{Markov23, Markov24, fisher2025rebelsifudustattenuationcurves} when the Universe was only $\sim 700$ Myr old. The presence of the UV bump at such early times challenges existing models of dust formation. Asymptotic giant branch (AGB) stars provide a likely origin for PAHs \citep{Latter_1991}, and the standard production channel of carbonaceous dust grains is thought to be through low mass ($\leq 3 M_\odot$) AGB stars reaching the end of their main-sequence lifetime on timescales exceeding $300$ Myr. 
If this is the dominant production channel of PAHs, the detection of the UV bump at $z\sim7$ implies that the onset of star formation occurred at $z\geq10$ \citep{Witstok_2023_Bump}.
For a galaxy where the onset of star formation occurs at $z=10$, low-mass AGB stars would begin dust production at $z\sim7$, so it is expected that supernovae (SNe) instead dominate dust production, with dust producing SNe II occurring $\sim10$ Myr after the onset of star formation (see \citealt{schneider_formation_2023} for a review).  Supporting the idea of early dust production, enhanced carbon-to-oxygen (C/O) ratios are seen in metal-poor stars in the MW, and have now been observed in GS-z12, a galaxy at $z=12.5$ \citep{deugenio2023}. While the chemical enrichment pattern of GS-z12 is inconsistent with pure SNe II yields, low-energy Population III SNe yields may explain the C/O lower limit \citep{Vanni_2023, deugenio2023}.
Furthermore, it has also been suggested that the slope of the attenuation curve flattens and the strength of the UV bump weakens with increasing redshift due to the grain size distribution changing, with larger dust grains at earlier epochs \citep{makiya_2022}, which could be due to SNe being the prominent channel of dust formation.

The dust attenuation curves of high redshift galaxies remained largely unconstrained until the launch of the \emph{James Webb Space Telescope} \citep[\emph{JWST};][]{mcelwain_23, Rigby_23}. With the Near-infrared Spectrograph \citep[NIRSpec;][]{jakobsen22, boker_23} onboard \emph{JWST}, we are now able to explore the dust attenuation curves of high redshift galaxies in more detail with successful detections of the UV bump \citep[][]{Witstok_2023_Bump, Markov23, Markov24} out to redshifts of 8, place constraints on the nebular attenuation curve of a galaxy at $z=4.41$ \citep{sanders2024aurora_nebular}, and explore the redshift evolution of dust attenuation curves \citep[e.g.,][]{Markov24}, where it has been suggested that the attenuation curve flattens with increasing redshift independent of $A_V$. 

Here, we combine NIRSpec Wide GTO observations with data from the \emph{JWST}/Near-infrared Camera \citep[NIRCam;][]{rieke2023} to investigate the presence of a UV bump in a galaxy at $z=7.11235$. This system is potentially undergoing a merger, allowing us to explore the implications of these findings for dust formation in the early Universe. 
This paper is structured as follows: in Section \ref{sec:observations} we discuss the observations used in this work, in Section \ref{sec:methods} we discuss our methods and analysis, and we place these into the context of galaxy and dust formation in Section \ref{sec: discussion}. Finally, our findings are summarized in Section \ref{sec:summary}.
Throughout this paper, we assume a standard cosmology of $H_0 = 70~{\rm kms}^{-1}{\rm Mpc}^{-1}$, $\Omega_m = 0.3$, and $\Omega_\Lambda = 0.7$ and a solar abundance of $12+\mathrm{log(O/H)} = 8.69$ \citep{Apslund_solar_2021}. All magnitudes are quoted in the AB magnitude system \citep{oke_gunn_1983}, and galaxy sizes refer to the half-light radius. 

\section{Observations}
\label{sec:observations}
GNWY-7379420231, at $z_\mathrm{prism}=7.11235$, was observed as part of the NIRSpec Wide Guaranteed Time Observations (GTO) Program \citep[henceforth referred to as Wide]{maseda_nirspec_2024}. The Wide survey covers the five extragalactic deep fields of the Cosmic Assembly Near-infrared Deep Extragalactic Legacy Survey \citep[CANDELS;][]{grogin_candels_2011, koekemoer_candels_2011} with 31 pointings, covering $\approx 320~$ arcmin$^2$ in $105$ hours.
As part of the high priority targets in the Wide survey, IRAC-excess sources from \citet{smit_high-precision_2015, roberts-borsani_z_2016} are observed, i.e. galaxies at $z=6-8$ with strong optical emission lines determined from \emph{Spitzer}/IRAC photometry. At the time of writing, the sample of IRAC-excess sources is made up of 23 galaxies, covering a redshift range of $z=5.66-7.65$. Through visual inspection of these sources, we identify the presence of a strong UV bump in GNWY-7379420231 only. We show a tentative UV bump detection in EGSZ-9135048459, at $z_\mathrm{prism}=6.74$, in Appendix \ref{egs appendix}. 

{\id} was originally identified as a Lyman break candidate at $z_\mathrm{phot}=8.29$ in \citet{Bouwens2015_UVLF}, and identified as a \emph{Spitzer}/IRAC-excess source in \citet{roberts-borsani_z_2016}. Subsequent Keck/MOSFIRE observations in \citet{Roberts_Borsani_2023_Lya} revealed Ly-$\alpha$ emission, confirming a spectroscopic redshift of $z_{\rm{Ly}\alpha}=7.10813\pm0.00019$. 

\subsection{Spectroscopic Observations} 
\label{sec:nirspec}
The low-resolution spectrum was obtained using PRISM/CLEAR (PRISM hereafter), covering a wavelength range of $0.6\mu\mathrm{m}-5.3\mu\mathrm{m}$ at a spectral resolution of $R~\approx~100$ (varying from $30-300$ for a uniformly illuminated shutter; \citealt{jakobsen22}). High-resolution spectra were obtained using the G235H and G395H gratings (and associated F170LP and F290LP filters), providing wavelength coverage of $1.66\mu\mathrm{m}-3.05\um$ and $2.78\um-5.14\um$, respectively, at a spectral resolution of $R~\approx~2700$.

For the PRISM spectrum, 1 exposure with 55 groups was taken, using the NRSIRS2RAPID read-out mode \citep{rauscher17} with a 3-point nodding pattern to cycle through the 3 shutter-slitlet per target. This results in an effective exposure time of $2451$s. The high resolution gratings were nodded between the two outer shutter positions. For the G235H grating, 1 exposure with 55 groups was taken for each of the two nodding positions (1634s total) and for the G395H grating 1 exposure with 60 groups was taken (1780s total). 

\subsection{NIRSpec Data Reduction}
\label{sec:reduction}
We use the same core reduction process as other NIRSpec Multi-Object Spectroscopy (MOS) GTO surveys \citep[e.g.,][]{Curtis-Lake_23, cameron_jades_2023, Carniani24, bunker2024, saxena24} developed by the ESA NIRSpec Science Operations Team (SOT) and GTO teams, as described in \citet{Carniani24}. Most of the pipeline uses the same algorithms as the official Space Telescope Science Institute (STScI) pipeline \citep{alves_de_oliveira_18, ferruit_22}, with small survey-specific modifications \citep{maseda_nirspec_2024}. We use a finer grid in wavelength with regular steps in the 2D rectification process. We also estimate path losses for each source by taking the relative intra-shutter position into account and assuming a point-like morphology, as in \citet{bunker2024, curti_jades_2024}. The Wide reduction differs from other GTO reductions in the sigma-clipping algorithm used to exclude outliers when creating the 1D spectrum, which does not work well with a low number of exposures as in Wide, and does not account for Poisson noise from bright pixels, which is typical for many Wide targets. Therefore, the reductions used in this work skip this step.

\subsection{NIRCam Imaging}
\label{sec:nircam}
We use NIRCam imaging of the Great Observatories Origins Deep Survey \citep{Giavalsico2004} North (GOODS-N) field, obtained as part of the \emph{JWST} Advanced Deep Extragalactic Survey \citep[JADES;][]{Eisenstein_JADES} from programme 1181 (PI: Eisenstein).  Nine filters in total are used within these observations: F090W, F115W, F150W, F200W, F277W, F335M, F356W, F410M and F444W.  The reduction of these images closely follows the method used in \citet{rieke2023}. 

\begin{figure*}
    \centering
    \includegraphics[width=0.9\linewidth]{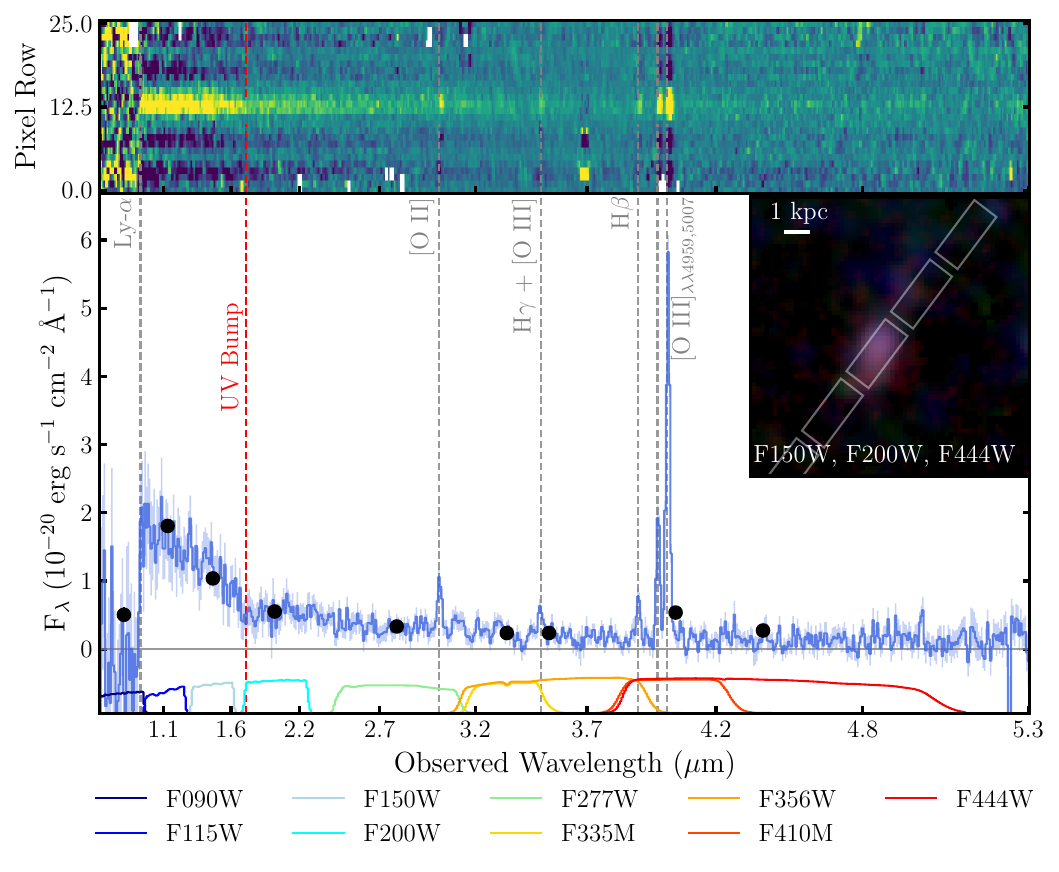}
    \caption{2D spectrum (top panel) and the 1D spectrum (bottom panel) of GNWY-7379420231 (solid blue line) with photometry overlaid (black points). The location of the 2175\AA\ UV bump is shown by the dashed red vertical line, and the locations of key emission lines are shown by dashed grey lines. The transmission curves for each photometric filter used are also shown. The inset panel shows a PSF-matched RGB image of GNWY-7379420231 (R: F444W, G: F200W, B: F150W) with the location of the NIRSpec slitlets outlined in solid white lines.}
    \label{fig:data}
\end{figure*}

\subsection{Aperture Correction}

We use NIRCam photometry extracted within Kron apertures with Kron parameter equal to 1.2 ($R_\mathrm{Kron, s}$) as high signal-to-noise estimates of flux in each filter band. For a galaxy with a Sérsic index of $n=1$, $R_\mathrm{Kron, s}$ corresponds to the half-light radius \citep{Graham_Driver_2005}. We use the Kron photometry extracted in apertures with Kron parameter equal to 2.5 ($R_\mathrm{Kron}$) to determine the ratio $R_\mathrm{Kron}/R_\mathrm{Kron,s}$ in the F444W band to estimate the amount of flux missed by the smaller apertures. We then correct all $R_\mathrm{Kron,s}$ fluxes by this factor to obtain our final flux values. 

In order to correct the spectrum to account for flux outside the slit, we first derive synthetic photometry of the NIRSpec spectrum using the NIRCam filter curves. We then multiply the spectrum by the median ratio of the scaled $R_\mathrm{Kron,s}$ photometry to the synthetic photometry, excluding F090W where the source is not detected. The 1D and 2D spectra are shown in Figure \ref{fig:data} with the NIRCam photometry and filter curves.

\section{Methods and Analysis}
\label{sec:methods}
\subsection{Spectroscopic Redshift}
We estimate the spectroscopic redshift from the PRISM spectrum with \texttt{msaexp}\footnote{\url{https://github.com/gbrammer/msaexp}} \citep{msaexp} using \textsc{eazy} templates \citep{brammer_2008_eazy}, as in \citet{maseda_nirspec_2024}. From this, we obtain a spectroscopic redshift of $z_\mathrm{prism}=7.11235$. We additionally estimate the [O\textsc{iii}]$\lambda5007${\AA} redshift from the G395H spectrum, obtaining a redshift of $z_\mathrm{[O\textsc{iii}], G395H}= 7.1082^{+0.000040}_{-0.000046}$. Although $z_\mathrm{[O\textsc{iii}], G395H}$ is closer to $z_{\rm{Ly}\alpha}$, we use $z_\mathrm{prism}$ for all analysis of the NIRSpec PRISM spectrum, and use $z_\mathrm{[O\textsc{iii}], G395H}$ for analysis of the high-resolution G395H spectrum.   

\subsection{UV Magnitudes and Slopes}
\label{sec:UV mag slope}
We convert the \emph{JWST} NIRCam F150W apparent magnitude ($m_{\rm UV}$) to give UV absolute magnitude ($M_{\rm UV}$), which is reported in Table \ref{tab:properties}.
We model the UV continuum slope of the galaxy considered here with a power law $F_\lambda \propto \lambda^{\beta}$. 
We measure this slope within the fitting windows defined in \citet{Calzetti_1994}, following the approach adopted in previous studies \citep[e.g.,][]{austin_epochs_2024, saxena_uv_slopes}. We exclude the windows redwards of 1833{\AA} to exclude the UV bump region and the C\textsc{iii}]$\lambda\lambda1907,1909$ doublet (C\textsc{iii}]). We include a fitting window at $2580-2640${\AA} to reduce the uncertainty on the fit. 

We use a Bayesian fitting procedure to fit the rest-frame UV continuum using a \textsc{python} implementation of the \textsc{multinest} nested sampling algorithm \citep{multinest}, \textsc{pymultinest} \citep{pymultinest}. We use a Gaussian prior distribution for the power-law index, centred on $\mu_\beta = -2$, with a standard deviation of $\sigma_\beta = 0.5$. This is motivated by studies such as \citet{austin_epochs_2024} and \citet{saxena_uv_slopes}, which find that the average UV continuum slope at $z\sim7$ is $\beta_\mathrm{UV}\sim -2$. We use a flat prior on the normalization at $\lambda_{\rm rest} = 1500$\AA\ (between 0 and twice the maximum flux value of the spectrum in the fitting regions). The best-fit value of $\beta$ is given by the 50th percentile (median) of the posterior distribution, with the 16th and 84th percentiles given as a $\pm 1 \sigma$ confidence range. The UV slope fit is shown in Figure \ref{fig:UV fit}. 

\subsection{UV Bump Identification}
\label{sec:bump identification}
To determine the robustness of the UV bump feature, we follow the method defined in \citet{Witstok_2023_Bump}. We fit power laws in four adjacent wavelength windows defined by \citet{Noll_2007}, with power-law indices $\gamma_1$ to $\gamma_4$. While the $\gamma_3$ region begins at 1920{\AA} in \citet{Noll_2007}, we exclude the region 1920\AA\ $< \lambda_\mathrm{emit} < $1950\AA\ to ensure we avoid contamination from the C\textsc{iii}] doublet. The parameter $\gamma_{34} \equiv \gamma_3 - \gamma_4$ is used to identify the presence of the absorption feature centred on 2175\AA, where a more prominent UV bump being present results in a more negative value of $
\gamma_{34}$ \citep{Noll_2009}. Prior to fitting the wavelength windows, the spectrum is smoothed with a running median filter of 15 pixels. The uncertainty of the running median is estimated with a bootstrapping procedure, where each of the 15 pixels is randomly perturbed according to their formal uncertainty for 100 iterations. 

We choose flat prior distributions for the power law indices within the range $-10 < \gamma_i < 1$, and normalise at the centre of each wavelength window, between 0 and twice the maximum value of the flux in each fitting region. The best-fit value of $\gamma_{34}$ is given by the median of the posterior distribution obtained by simultaneously fitting $\gamma_3$ and $\gamma_4$, with the 16th and 84th percentiles given as a $\pm1\sigma$ confidence range.

\subsection{UV Bump Fitting}
\label{sec:bump fitting}
As in \citet{Witstok_2023_Bump}, we parameterise the UV bump profile by defining the excess attenuation as $A_{\lambda, \text { bump }}=-2.5 \log _{10}\left(F_\lambda / F_{\lambda, \text { cont }}\right)$, 
where $F_\lambda$ is the observed flux, and $F_{\lambda, \text { cont }}$ is the UV continuum slope, without a UV bump \citep{Shivaei_2022}.  We use the running median and its corresponding uncertainty (see Section \ref{sec:bump identification}) to compute the significance of the negative flux excess of the spectrum with respect to the power-law UV slope ($\beta_{\mathrm{UV}}$), determined in Section \ref{sec:UV mag slope}. 

\begin{figure*}
    \centering
    \includegraphics[width=0.9\linewidth]{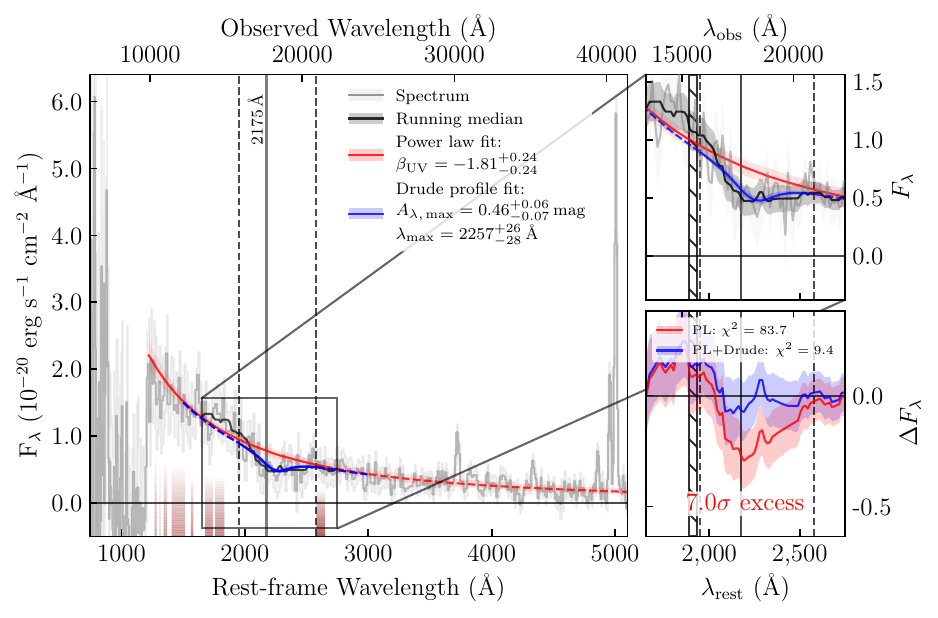}
    \caption{Spectrum of GNWY-7379420231 (grey solid line) with a power-law fit to the UV continuum (red solid line). The dark red shading represents the UV slope fitting windows. The zoom in panel of the region around 2175\AA\ shows the running median, indicated by a solid black line. This represents the attenuated stellar continuum, and shows a localized absorption feature. The Drude profile fit is shown by the solid blue line, within the fitting window indicated by the vertical dashed lines. The hatched region shows the location of the C\textsc{iii} doublet. The bottom right panel shows the residuals of the power-law fit (PL) and the combined power-law and Drude profile fit (PL+Drude). The power-law fit alone has a $7.0\sigma$ negative flux excess, with the PL+Drude model showing a significantly better fit, as supported by the BIC values. }
    \label{fig:UV fit}
\end{figure*}

We use the \textsc{multinest} nested sampling algorithm to fit the excess attenuation $A_{\lambda, \text { bump }}$ with a Drude profile \citep{Fitzpatrick_1986}. The Drude profile is characterized by the parameters $A_{\lambda, \max }$, the peak amplitude of the bump; $\lambda_{\mathrm{max}}$, the rest-frame central wavelength of the bump; and $\gamma$, which is related to the full width at half maximum (FWHM) as FWHM$=\gamma\lambda^{2}_{\mathrm{max}}$. The Drude profile is then defined as
\begin{equation}
A_{\lambda, \text { bump }}=A_{\lambda, \max } \frac{\gamma^2 / \lambda^2}{\left(1 / \lambda^2-1 / \lambda_{\max }^2\right)^2+\gamma^2 / \lambda^2}.
\end{equation}
We fix $\gamma=250$\AA$ /(2175 $\AA$)^2$ which corresponds to FWHM$= 250$\AA\ if $\lambda_\mathrm{max} = 2175${\AA}, in agreement with findings for $z\sim2$ star-forming galaxies \citep{Noll_2009, Shivaei_2022}. We note that allowing the FWHM to vary does not affect the other parameters. We carry out the fitting procedure in a region of 1950\AA\ $< \lambda_\mathrm{emit} < $ 2580\AA, which includes the $\gamma_3$ and $\gamma_4$ regions, but excludes the C\textsc{iii}] doublet. 
As in \citet{Witstok_2023_Bump}, we use a gamma distribution with shape parameter $a =1$ and scale $\theta = 0.2$ as a prior for bump amplitude, $A_{\lambda, \mathrm{max}}$. This prior is chosen as it favours lower amplitudes, consistent with our expectations for high redshift galaxies \citep[e.g.,][]{Markov24}, although a flat prior gives comparable results. We use a flat prior for the central wavelength in the range 2100\AA\ $< \lambda_\mathrm{max} <~$2300\AA.  The UV bump fit is shown in Figure \ref{fig:UV fit}, with best-fit values listed in Table \ref{tab:properties}. We compare this model to a simple power-law using the Bayesian Information Criterion (BIC). The significantly lower BIC value for the UV bump model ($\Delta\rm{BIC}=66.7$) indicates that this model is preferred over a simple power-law model. 

\subsection{SED Fitting}
\label{sec:sed fitting}
We model the SED of GNWY-7379420231 using \textsc{bagpipes} \citep[Bayesian Analysis of Galaxies for Physical Inference and Parameter EStimation;][]{Carnall_2018, Carnall_2019}, fitting both the photometry and spectroscopy simultaneously. We use Binary Population and Spectral Synthesis (\textsc{bpass}) v2.2.1 models \citep{BPASS}, with models that include binary stars. We use the default \textsc{bpass} initial mass function \texttt{"135$\_$300"} (IMF; stellar mass from $0.1\rm{M}_\odot$ to $300\rm{M}_\odot$ and a slope of $-2.35$ for $M > 0.5\rm{M}_\odot$). We mask the region around Ly$\alpha$ (1200~\AA\ $< \lambda_{\rm{emit}}<~$1250~\AA) to mitigate any potential effects from Lyman-$\alpha$ damping-wing absorption. 
We also mask the region 4900~\AA\ $< \lambda_{\rm{emit}}<~$5100~\AA \ in the spectrum due to  small discrepancies between the [O\textsc{iii}] equivalent widths (EW) in the spectroscopy and photometry. While \textsc{bagpipes} can account for small wavelength-dependent variations between the spectroscopy and photometry, it is not flexible enough to solve for emission line discrepancies.
We have verified that the SED fitting results do not change significantly if the spectrum is left unmasked.

We assume a non-parametric star formation history (SFH) from \citet{Leja19}, which fits the star formation rates in fixed time bins, where $\Delta$logSFR between bins is linked by a Student's t-distribution. As in \citet{Tacchella22}, we fit a `continuity' model, with $\sigma = 0.3$ and $\nu = 2$, which is weighted against rapid transitions in star formation rate, and a `bursty continuity' model with $\sigma = 1$ and $\nu = 2$, which allows more variation in star formation rate (SFR) (i.e., a more bursty star formation history). We fix the redshift of the galaxy to $z_\mathrm{prism} = 7.11235$, and fit the SFH in 6 bins of lookback time $t$, with the first bins fixed to $0~\rm{Myr} < t < 3~\rm{Myr}$ and $3~\rm{Myr} < t < 10~\rm{Myr}$, motivated by observational evidence for bursty SFHs seen at high-redshift \citep[e.g.,][]{looser_jades_2023, Boyett_2024, endlsey_2024_burstiness} and following the same approach adopted by \citet{whitler_ages_2023}.
The remaining four bins are equally log-spaced in lookback time until $t(z=20)$. We note that the inferred SFH is largely insensitive to the number of bins used as long as $N_\mathrm{bins}\geq4$ \citep{Leja19}. We favour the `bursty continuity' model as this flexible SFH accommodates both stochastic star formation and underlying older stellar populations. It should be noted that this model still allows smooth or declining SFHs if favoured during the fitting \citep[see][]{harvey_epochs_2024}. We reduce the star formation timescale to $10~$Myr to account for the increased specific star formation rate (sSFR), compared to galaxies at lower redshift.

The allowed total stellar mass formed and stellar metallicities are allowed to vary uniformly between $10^5~\rm{M}_\odot < M_\star < 10^{12} ~\rm{M}_\odot$, and $0<Z_*<0.5~ \mathrm{Z}_{\odot}$, respectively. Nebular emission is included using a grid of \textsc{cloudy} \citep{Cloudy} models, parameterized by the ionization parameter $(-3 < \rm{log}_{10} U < -0.5)$, which \textsc{bagpipes} computes self-consistently. We use the \citet{Salim2018} dust attenuation curve, which parameterises the dust curve shape with a power-law deviation $\delta$ from the \citet{Calzetti00} model ($\delta = 0$ for the Calzetti curve) and includes a Drude profile to model the 2175\AA\ bump. The strength of the bump is given by the amplitude $B$, in units of $A_{\text {bump }} / E(B-V)$ where $A_\text{{bump}}$ is the extra attenuation at 2175\AA. We keep the central wavelength and width of the bump fixed at 2175\AA\ and 250\AA, respectively. We use uniform priors for $\delta$ ($-0.5<\delta<0.2$, adapted from \citealt{kriek_conroy_2013}) and $B$ $\left(0< B <10\right)$. Following \citet{Witstok_2023_Bump}, we use a Gaussian prior on the V-band dust attenuation with $\mu_{A_V} = 0.15$ mag, $\sigma_{A_V} = 0.15$ mag and attenuation limited to $0 < A_V < 7$ mag, fixing the fraction of attenuation arising from stellar birth clouds to $60\%$, with the remaining $40\%$ coming from the diffuse ISM \citep{chevallard_2019}. We also assume a log-prior on the velocity dispersion in the range $1-1000$ km~s$^{-1}$. 
Finally, we assume that the spectrum follows the PRISM resolution curve based on a point-source morphology, using the resolution curve of an idealized point source generated with \texttt{msafit}, as described in \citet[][Appendix A]{de_graaff_lsf}.

\begin{figure*}
    \centering
    \includegraphics[width=0.99\textwidth]{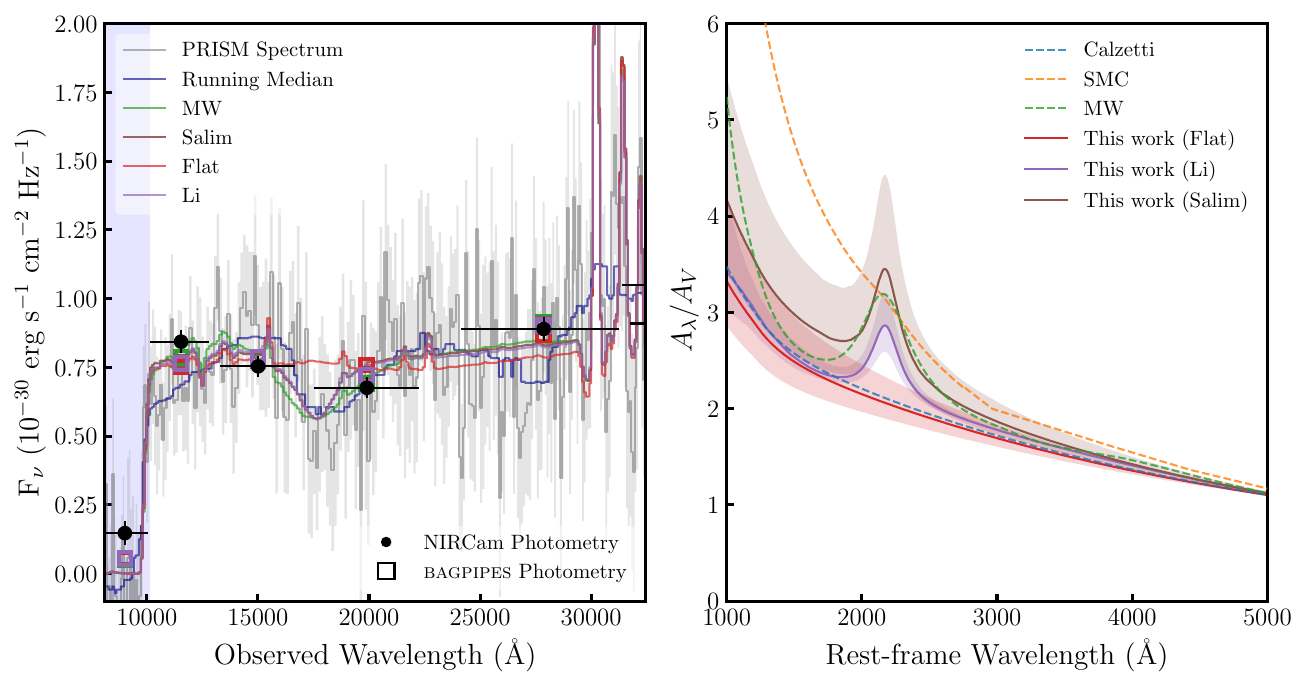}
    \caption{Left: Posterior spectra obtained from combined \textsc{bagpipes} fitting of the photometry and spectroscopy. The observed spectrum and associated errors are shown in grey, while the running median is shown in dark blue. The x error bars represent the width of the filter at $50\%$ of the maximum transmission. The blue shading shows the spectral regions masked in the fitting process.
    Right: The best-fit dust attenuation curve obtained from our \textsc{bagpipes} fit with the Salim dust curve is shown by the solid brown line, with the brown shaded region showing the $1\sigma$ uncertainty. The best-fit Li model and $1\sigma$ uncertainty is shown by the solid purple line, and purple shading, and the red solid line represents the best-fit dust attenuation curve from the flat Salim fit. We show the Calzetti, MW, and SMC curves for comparison, in the blue, orange, and green dashed lines, respectively.} 
    \label{fig:dust}
\end{figure*}

We carry out further \textsc{bagpipes} fits, varying only the assumed dust attenuation curve. We perform fits with the Milky Way dust curve, the Salim dust law with $B$ fixed as 0 (henceforth referred to as the flat Salim curve), allowing only the slope ($\delta$) to deviate from the Calzetti curve, and the \citet{Li_2008} analytical expression for the dust attenuation law, used in \citet{Markov23, Markov24}. This dust parameterization has the benefit of allowing SED fitting to be carried out without assuming the prior shape of the dust curve, but has the disadvantage of having four free parameters, thus should only be used with spectroscopic data, or photometric surveys with a sufficiently large number of photometric bands \citep{Markov23}. We modify the expression to allow for a variable width of the Drude profile characterising the UV bump, and as such the dust curve, normalized to the attenuation at $0.55\mu$m ($A_V$), is given by 
\begin{equation}
\begin{aligned}
\label{eq:A_V_curve_Li_2008}
A_\lambda / A_V= & \frac{c_1}{(\lambda / 0.08)^{c_2}+(0.08 / \lambda)^{c_2}+c_3} \\
& +\frac{233\left[1-c_1 /\left(6.88^{c_2}+0.145^{c_2}+c_3\right)-c_4 / x\right]}{(\lambda / 0.046)^2+(0.046 / \lambda)^2+90} \\
& +\frac{c_4}{(\lambda / 0.2175)^2+(0.2175 / \lambda)^2 + \left[(w/0.2175)^2-2 \right]},
\end{aligned}
\end{equation}
where $c_1$, $c_2$, $c_3$, $c_4$ are dimensionless parameters, $\lambda$ is the wavelength in $\mu$m, $x = (6.55 + \left[(w/0.2175)^2-2 \right])$, and $w$ is the width of the UV bump in $\mu$m. The three terms of \cref{eq:A_V_curve_Li_2008} describe the far-ultraviolet (FUV) rise, the optical and near-infrared (NIR) attenuation, and the $2175$\AA\ UV bump. We use priors for $c_1 - c_4$ adapted from those given in \citet{Markov24}, requiring $c_4 \geq 0$, and set $w=0.0250\mu$m.
We present the best-fit values obtained with each dust curve, along with $1\sigma$ errors in Table \ref{tab:bagpipes_comp}. 
We summarise the priors used in Table \ref{tab:priors}.
The posterior spectra and dust curves are shown in Figure \ref{fig:dust}. We show the full posterior spectra with residuals in Figure \ref{fig:bagpipes_models}. 

We find that fitting the photometry or spectrum alone recovers similar parameters for the UV bump (see Figure \ref{fig:bagpipes_phot_only}).

{\renewcommand{\arraystretch}{1.5}
\begin{table}
    \caption{The properties of GNWY-7379420231. Error bars represent $1\sigma$ uncertainties. }
    \centering
    \begin{tabular}{l|c}
        Parameter & Value  \\ \hline
        Wide ID & 2008001576 \\
        R.A. & 12:37:37.941\\
        Dec. & +62:20:22.850\\
        $z_\mathrm{prism}$ & 7.11235\\
        $z_\mathrm{[O\textsc{iii}], G395H}$ & $7.1082^{+0.000040}_{-0.000046}$ \\
        $M_\mathrm{UV}$ (mag) & -20.29\\
        $\beta_\mathrm{UV}$ & $-1.81\pm0.24$\\
        $\gamma_{34}$ & $-6.58^{+1.85}_{-1.69}$\\ 
        $A_{\lambda,\mathrm{max}}$ (mag) & $0.46^{+0.06}_{-0.07}$\\
        $\lambda_\mathrm{max}$ (\AA) & $2257^{+26}_{-28}$\\
        $[\mathrm{O}\textsc{iii}]$+H$\beta$ EW$_0$ (\AA) &  $1110\pm183$\\
        $\log_{10}(\mathrm{O}{32})$ & $0.39\pm0.19$\\
        $\log_{10}(\mathrm{R}{23})$ & $1.23\pm0.24$\\ 
        $\log_{10}(\mathrm{R}2)$ & $0.59\pm0.29$\\
        $\log_{10}(\mathrm{R}3)$ & $0.98\pm0.25$\\
        $\hat{R}$ & $1.14\pm0.26$\\
        $Z_{\mathrm{neb}} ~(Z_\odot)$ & $0.30^{+0.10}_{-0.09}$\\
        $\log_{10}(n_e)$ (cm$^{-1}$) & $3.12^{+0.83}_{-0.47}$\\
        $\log_{10}(M_\mathrm{dyn}/M_\odot)$ & $9.35\pm0.43$\\
        \hline
        \end{tabular}
    \label{tab:properties}
\end{table}}

\subsection{Morphology}
\label{sec:morph}
We perform surface brightness fitting with \textsc{galfit} version 3.0.5 \citep{Galfit1, Galfit2} to investigate the morphology of our source. \textsc{galfit} convolves the galaxy surface brightness profile with a point spread function (PSF), and uses the Levenberg-Marquardt algorithm to minimise the $\chi^2$ of the fit. 

We create PSF matched images using \textsc{aperpy} \footnote{\texttt{aperpy} is available though Github (\url{https://github.com/astrowhit/aperpy}) and Zenodo (\url{https://doi.org/10.5281/zenodo.8280270}).} \citep{Weaver2023_aperpy}, which creates empirical PSFs and makes use of \textsc{pypher} \citep{Boucaud_2016} to create PSF matched kernels. We create a PSF matched stack of the filters in the short wavelength (SW) channel where the source is detected (F115W, F150W, F200W) to determine the components required for fitting. 

\begin{figure}
    \centering
    \includegraphics[width=0.99\columnwidth]{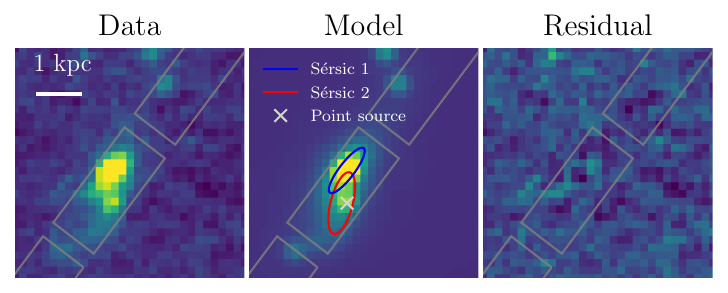}
    \caption{Left: the stacked data image provided as input to \textsc{galfit}. Middle: The \textsc{galfit} model image. The two Sérsic profiles are shown by the blue and red ellipses, with the point source indicated by the grey cross. Right: The residual image, created by subtracting the model image from the data image. The data and model images are linearly scaled between $-3\text{--}20\sigma$ of the data image background, and the residual image is scaled between $-3\text{--}10\sigma$ of the data image background, for clarity. The location of the NIRSpec slitlets are overplotted in grey.}
    \label{fig:galfit_stack}
\end{figure}

We fit the source using two Sérsic components and a point source. 
The Sérsic profile has the form
\begin{equation}
    I(R)=I_e \exp \left\{-b_n\left[\left(\frac{R}{R_e}\right)^{1 / n}-1\right]\right\},
    \label{eqn:Sérsic}
\end{equation}
where $I(R)$ is the intensity at a distance $R$ from the centre of the galaxy, $R_{e}$ is the half-light radius of galaxy, $I_e$ is the intensity at the half-light radius, $n$ is the Sérsic index \citep{Sersic, ciotti_1991, caon_1993}, and $b_{n}$ can be approximated as $b_n\approx 2 n-\frac{1}{3}+\frac{4}{405 n}+\frac{46}{25515 n^2}$ \citep{ciotti1999}.  We allow $R_e$, magnitude, axis ratio ($b/a$) and position angle to vary freely,  allow the Sérsic index to vary between $0 < n < 10$, and allow the source position to vary within $\pm3$ pixels of the input location in both the $x$ and $y$ direction. We find that one component has a very low Sérsic index, which we fix as $n=0.2$.  The best-fit model for the SW stack is shown in Figure \ref{fig:galfit_stack}. We then fit each band individually, keeping all parameters fixed to the best-fit values while allowing only the magnitude to vary. The best-fit models are shown in Figure \ref{fig:galfit_bands}, where the point source component is particularly visible in the SW channels.

The best-fit model is made up of a main, brighter component (Sérsic 1), with a `tail', made up of the secondary Sérsic component (Sérsic 2) and a point source. The bright region (Sérsic 1) is best-fit by a Sérsic profile with $n=6.04\pm2.27$, 
and has a half-light radius of $R_e = 569\pm170$ pc. The fainter Sérsic component (Sérsic 2) is the component that is fit with a fixed Sérsic index of $n=0.20$, and has a half-light radius of $R_e = 648\pm80$ pc. Errors are the $1\sigma$ uncertainties derived by \textsc{galfit}. 

\begin{figure*}
    \centering
    \includegraphics[width=0.99\textwidth]{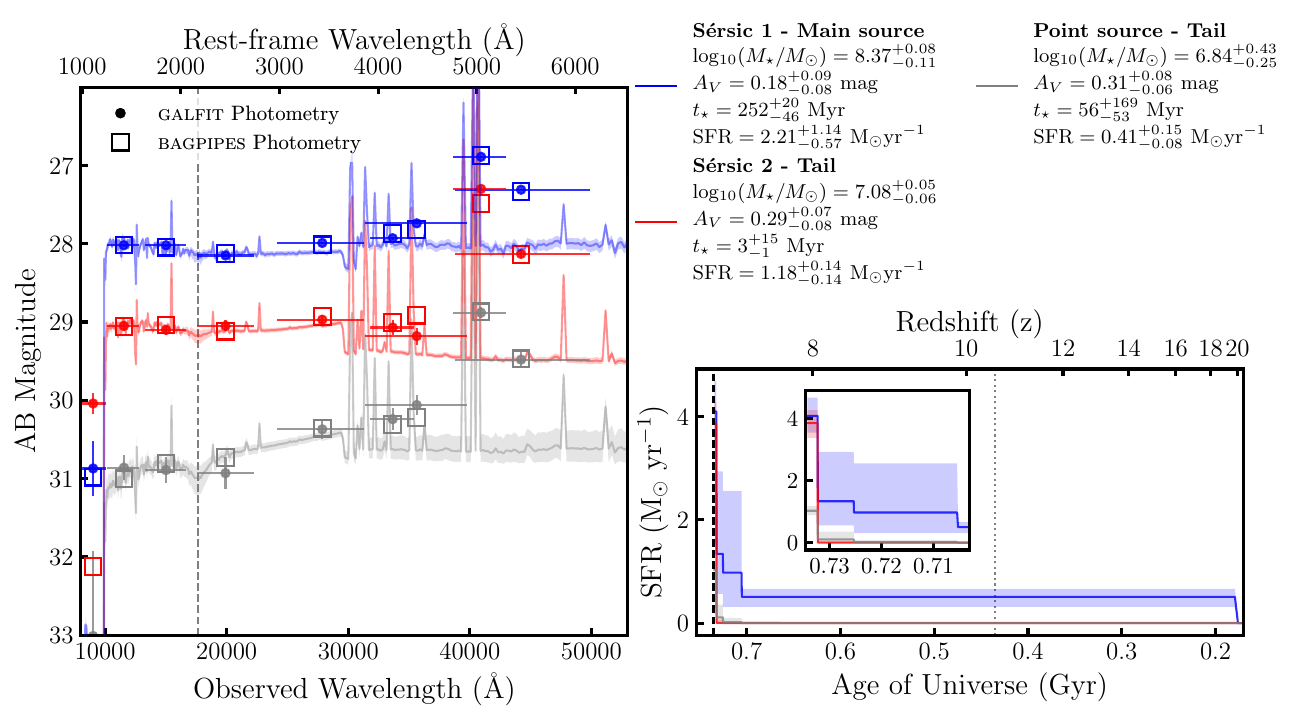}
    \caption{Left: Posterior spectrum for each component obtained through \textsc{bagpipes} SED fitting with $1\sigma$ errors. The \textsc{galfit} photometry is shown by the solid circles, with the \textsc{bagpipes} photometry shown by the open squares. The x error bars show the filter width at $50\%$ of the maximum transmission. Right: The posterior SFH for each component. The grey dotted line indicates where stars would have time (300 Myr) to evolve off the main sequence into an AGB star. The inset panel shows a zoom-in of the three most recent bins of the posterior SFH.}
    \label{fig:galfit_bagpipes}
\end{figure*}

We carry out photometric SED fitting for each of these three components, with the magnitudes obtained from the \textsc{galfit} fit in each band. 
As in Section \ref{sec:sed fitting}, we use \textsc{bpass} v2.2.1 models with the default \textsc{bpass} IMF.  We keep the redshift of the source fixed and use a `bursty continuity' model of star formation history, with bins of $0-3$ Myr, $3-10$ Myr, $10-30$ Myr and $30-t(z=20)$ Myr. The ionization parameter and metallicity are fixed to the best fit values obtained with the Salim dust law in Section \ref{sec:sed fitting}. We use the Salim dust law where the prior on $B$ is a truncated Gaussian prior with $\mu_B = 0$, $\sigma_B = 2$, and $0 < B < 10$.
The best-fit spectra and photometry are shown in Figure \ref{fig:galfit_bagpipes}, along with the posterior SFHs. 

Finally, we use \textsc{statmorph} \citep{statmorph} to measure the Gini-$M_{20}$ statistics, which can be used to quantify galaxy morphology \citep{Lotz2004, Lotz2008}. The Gini coefficient ($G$) is a statistic that is commonly used in economics to measure wealth distribution in human populations, but was first used by \citet{Abraham2003} to provide a quantitative measure of the distribution of light within a galaxy. $G$ is determined from the distribution of the absolute flux values: 
\begin{equation}
G=\frac{1}{\overline{|X|} n(n-1)} \sum_i^n(2 i-n-1)\left|X_i\right| ,
\end{equation}
where $\overline{|X|}$ is the mean of the absolute values $|X_i|$ \citep{Lotz2004, statmorph}. The $M_{20}$ statistic is defined as the normalized second order moment of the brightest 20 per cent of the galaxy's flux. $M_{20}$ traces the spatial extent of the brightest pixels in a galaxy, and is defined as 
\begin{equation}
M_{20} \equiv \log_{10}\left(\frac{\sum_i M_i}{M_{\mathrm{tot}}}\right), \text { while } \sum_i f_i<0.2 f_{\mathrm{tot}},
\end{equation}
where $M_{\rm{tot}}$ is defined as 
\begin{equation}
M_{\mathrm{tot}}=\sum_i^n M_i=\sum_i^n f_i\left[\left(x_i-x_c\right)^2+\left(y_i-y_c\right)^2\right],
\end{equation}
where $x_c$, $y_c$ is the galaxy's centre, such that $M_{\rm{tot}}$ is minimized \citep{Lotz2004, Lotz2008}.
We show the location of {\id} on the Gini-$M_{20}$ parameter space in Figure \ref{fig:gini_m20}, adopting the classifications from \citet{Lotz2008}. Gini and $M_{20}$ can be used to determine whether a galaxy is a merger, if the following criterion is met: 
\begin{equation}
    G > -0.14M_{20} + 0.33.
\end{equation}
Based on these diagnostics, {\id} falls within the merger region of the parameter space. The stellar mass ratio of 12:1 between Sérsic 1 and the two other components, derived from \textsc{bagpipes} spectral fitting, classifies this system as a minor merger. We also see evidence for a minor merger in the G395H 2D spectrum of the [O\textsc{iii}] doublet, with a velocity offset of $\sim180~$kms$^{-1}$, shown in Figure \ref{fig:vel offset}.

\subsection{Resolved SED Fitting}
\label{sec:resolved}

We create PSF matched images following the same method used in Section \ref{sec:morph}. We choose to match to the F444W mosaic as this has the broadest PSF of our filters. We create an inverse variance weighted stacked image of our source and use \textsc{vorbin} \citep{Cappellari2003} to perform adaptive spatial noise binning to create bins with a target signal to noise ratio (SNR) of 25. We extract photometry in each NIRCam filter for each bin, and model the SED of each bin using \textsc{bagpipes}. 

\begin{figure*}
    \centering
    \includegraphics[width=0.95\linewidth]{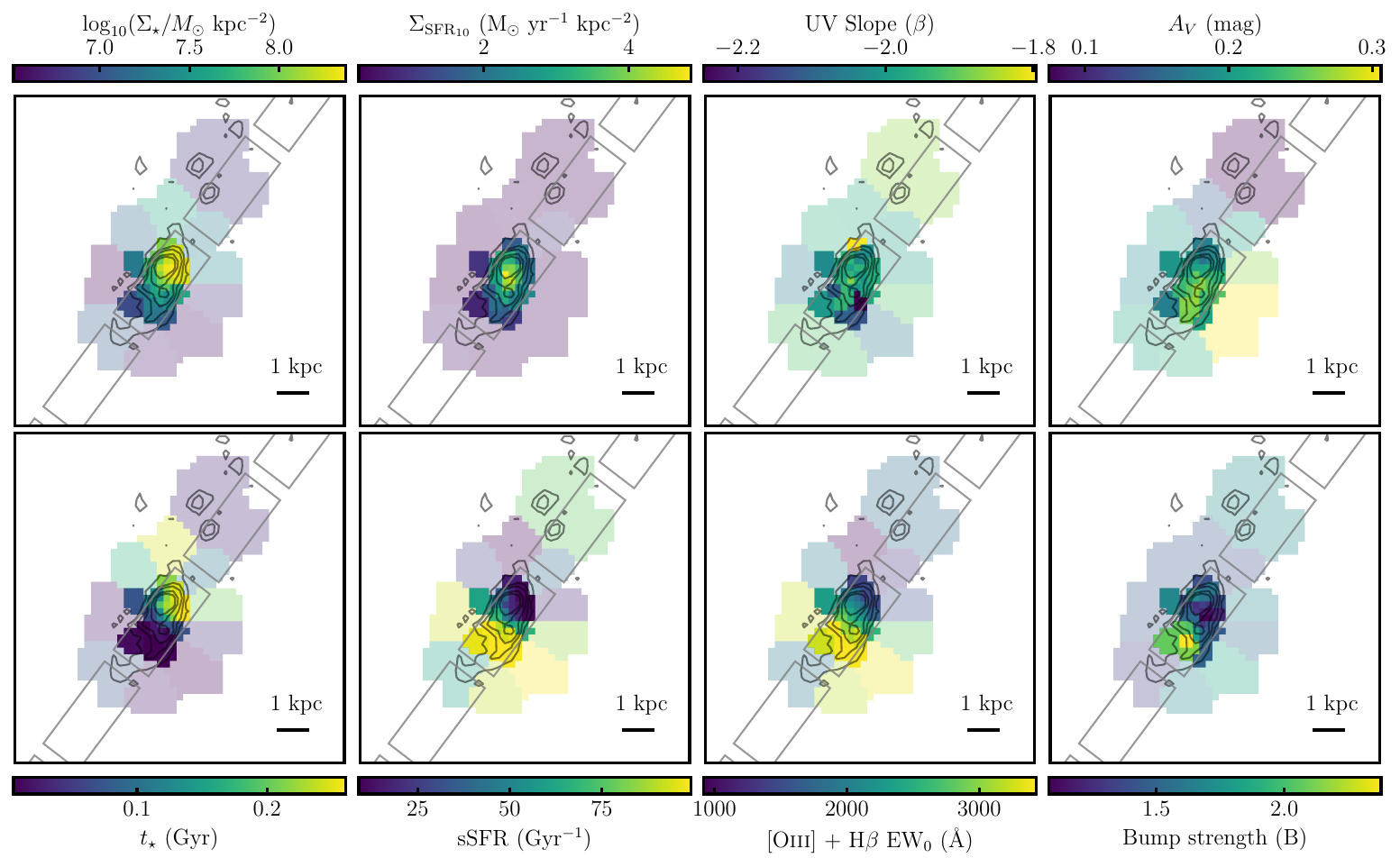}
    \caption{Maps of GNWY-7379420231. Top, from left to right: stellar mass surface density ($\Sigma_\star$), star formation rate surface density ($\Sigma_{\mathrm{SFR}_{10}}$), UV continuum slope, and V-band dust attenuation. Bottom, from left to right: mass-weighted age, specific star formation rate, [O\textsc{iii}]+H$\beta$ rest-frame equivalent width, and the UV bump strength ($B$) obtained from the Salim dust law (see Section \ref{sec:sed fitting}). The black lines show the 3, 6, 9, 15, and 21$\sigma$ contours of the SW stack. The location of the NIRSpec slitlets are overlaid in grey. The fainter bins are those where the photometry has at least one band with SNR$<5$ (excluding F090W).}
    \label{fig:resolved}
\end{figure*}

We apply the same \textsc{bagpipes} fitting procedure described in Section \ref{sec:morph}. 
We again adopt the Salim dust law with a truncated Gaussian prior on $B$ ($\mu_B = 0$, $\sigma_B = 2$, $0 < B < 10$) to allow us to trace the location of the UV bump.
Using the 50th percentile of the posterior \textsc{bagpipes} distributions, we create 2D maps of the physical properties of our source, as shown in Figure \ref{fig:resolved}. We measure the [O\textsc{iii}]+H$\beta$ rest-frame equivalent width from the 50th percentile of the posterior spectrum generated in the fitting process, and measure the UV $\beta$ slopes using the same fitting windows as in Section \ref{sec:UV mag slope}.

\subsection{Emission Line Measurements}
\label{sec:emission lines}
We perform emission line fitting that accounts for both the line spread function (LSF) broadening and its undersampling by the NIRSpec detectors. It is important that this is accounted for, as fitting a Gaussian to an undersampled line could severely over or underestimate the line flux \citep{de_graaff_RUBIES}. To address this, we create Gaussian models on an oversampled grid and convolve them with the LSF of an idealized point source from \citet{de_graaff_lsf}. We use the \textsc{multinest} nested sampling algorithm to fit the emission lines in both the PRISM and G395H spectra.

We fit the following emission lines in the PRISM spectrum: [O\textsc{ii}]$\lambda3727$, [O\textsc{ii}]$\lambda3729$, H$\beta$, [O\textsc{iii}]$\lambda4959$ and [O\textsc{iii}]$\lambda5007$. Due to the resolution of the PRISM spectrum, we fit the blended [O\textsc{ii}]$\lambda3727$+ [O\textsc{ii}]$\lambda3729$ emission lines as a single Gaussian. To reduce the number of free parameters, we fix the central wavelength of each Gaussian profile, and fix the flux ratio of the [O\textsc{iii}]$\lambda\lambda4959,5007$ doublet to the theoretical value of 2.98. 
As H$\alpha$ falls outside of the PRISM wavelength range, we instead estimate the dust attenuation from the value of $A_V$ obtained from SED fitting assuming the Salim dust law. We note this is in good agreement with that obtained from the \citet{meurer99} relation. We convert this to a nebular $A_V$ following \citet{reddy_mosdef_2020} and correct for dust attenuation assuming the \citet{Cardelli_1989} attenuation curve.

The dust corrected emission line fluxes are reported in Table \ref{tab:line fluxes} as the median of the posterior flux distribution, with errors given as the semi-difference of the 16th-84th percentiles of the posterior distribution.
As [O\textsc{iii}]$\lambda4363$ is blended with H$\gamma$ in the PRISM spectrum and falls within the chip-gap in the G395H spectrum, we are unable to derive the metallicity of {\id} using the direct-$T_e$ method \citep[e.g.][]{sanders_direct_2024}. Therefore, we derive the oxygen abundance ($12+\log$(O/H)) from a range of emission line ratios: 

\begin{equation}
\mathrm{R}2 =\log \left(\frac{[\mathrm{O\textsc{ii}}] \lambda \lambda 3727,3729}{\mathrm{H} \beta}\right)
\end{equation}
\begin{equation}
\mathrm{O}32 =\log \left(\frac{[\mathrm{O\textsc{iii}}] \lambda 5007}{[\mathrm{O\textsc{ii}}] \lambda \lambda 3727,3729}\right)
\end{equation}
\begin{equation}
\mathrm{R}3 =\log \left(\frac{[\mathrm{O\textsc{iii}}] \lambda 5007}{\mathrm{H} \beta}\right)
\end{equation}
\begin{equation}
\mathrm{R}23 =\log \left(\frac{[\mathrm{O\textsc{ii}}] \lambda \lambda 3727,3729+[\mathrm{O\textsc{iii}}] \lambda \lambda 4959,5007}{\mathrm{H} \beta}\right)
\end{equation}
\begin{equation}
\hat{R} =0.47 \times \mathrm{R}2+0.88 \times \mathrm{R}3.
\end{equation}

Figure \ref{fig:O32_R23} shows the dust corrected O32 and R23 emission line ratios. 
The dust corrected emission line ratios are reported in full in Table \ref{tab:properties}, along with the rest-frame equivalent width of the [O\textsc{iii}]$\lambda\lambda4959,5007$ doublet and H$\beta$ line. These equivalent widths are obtained using continuum fits from \textsc{ppxf} \citep{cappellari_improving_2017, cappellari_full_2023}, which will be described in Ormerod et al. (in prep).
We combine the information from the emission line ratios and use the calibrations from \citet{Curti_24_metallicity} to calculate the gas-phase metallicity ($Z_\mathrm{neb}$) in units of solar metallicity ($Z_\odot$), which is given in Table \ref{tab:properties}.

Using the $R\sim2700$ G395H grating, we measure the [O\textsc{ii}]$\lambda\lambda3727$,$3729$ and [O\textsc{iii}]$\lambda\lambda4959$,$5007$ emission lines. We first measure the [O\textsc{iii}] doublet with tied line widths. We use the median width of the posterior distribution as a fixed width when fitting the [O\textsc{ii}] doublet, where we also keep the central wavelengths fixed. The [O\textsc{ii}] fit is shown in Figure \ref{fig:electron_density}.

\subsection{Electron Density Measurement}
\label{sec:electron density}

Electron densities in H\textsc{ii} regions are crucial for characterising the ISM, as along with ISM pressure, they govern the emission from H\textsc{ii} regions \citep[e.g.][]{Kewley2019a, Isobe2023, abdurrouf2024}. 
To derive the electron density, we utilise the density-sensitive [O\textsc{ii}]$\lambda3726\rm{,}\lambda\lambda3729$ emission line ratio \citep{Kewley2019b}, measured from the G395H grating.

We then use \textsc{pyneb} \citep{pyneb_2015} to determine $n_e$ from the [O\textsc{ii}] line ratio. 
We adopt an electron temperature of $T_e = 10000$K \citep[e.g.,][]{Glass_electron_density}, however assuming an electron temperature of $15000$K or $20000$K has minimal impact on our derived electron density, consistent with the findings by \citet{topping_aurora_25}. 
From this, we obtain a value of $\log_{10}(n_e)$ cm$^{-1}$ = $3.12^{+0.83}_{-0.47}$, in agreement with the median value determined in \citet{Isobe2023} for $z\sim7-9$ galaxies.  

\subsection{Dynamical Mass}
\label{sec:dynamical mass}

We follow the method described in \citet{Kohandel2019} to estimate the dynamical mass of our source, which we summarise here. Assuming a rotating disk geometry with radius $R$, the dynamical mass can be estimated as 
\begin{equation}
\label{eqn:m_dyn}
    M_\mathrm{dyn} = \frac{v^2_c R}{G},
\end{equation}
where $v_c$ can be estimated from the FWHM of the [O\textsc{iii}]$\lambda5007$ emission line using
\begin{equation}
\label{eqn:fwhm}
    \mathrm{FWHM} = \gamma v_c \mathrm{sin}\theta,
\end{equation}
where $\gamma$ is a factor dependent on geometry, line profile, and turbulence. As in \citet{Capak2015}, we estimate $\gamma=1.32$ and additionally include a systematic error on $\gamma$ of $20\%$. 
It is important to note that both \citet{Capak2015} and \citet{Kohandel2019} derive dynamical masses using the $[$C\textsc{ii}$]~158\mu$m emission line, rather than the $[$O\textsc{iii}$]\lambda5007$ line employed in this analysis. Dynamical masses derived from the $[$O\textsc{iii}$]\lambda5007$ emission line could be overestimated compared to those derived from cold gas tracers \citep{kohandel_dynamically_2024}.
We determine the inclination ($\theta$) from the axis ratio of Sérsic 1 measured in Section \ref{sec:morph} using the \citet{Hubble1926} equation
\begin{equation}
    \cos ^2(\theta)=\frac{(b / a)^2-(b / a)_{\min }^2}{1-(b / a)_{\min }^2},
\end{equation}
where $(b/a)_\mathrm{min}=0.15$ \citep[e.g.,][]{Guthrie1992, Yuan2004, Sargent2010, Leslie2018}. Using Equations \ref{eqn:m_dyn} and \ref{eqn:fwhm}, the general expression for the dynamical mass is:
\begin{equation}
\label{eqn: m_dyn general}
\mathrm{M}_{\mathrm{dyn}}=2.35 \times 10^9 \mathrm{M}_{\odot}\left(\frac{1}{\gamma^2 \sin ^2 \theta}\right)\left(\frac{\mathrm{FWHM}}{100 \mathrm{~km} \mathrm{~s}^{-1}}\right)^2\left(\frac{\mathrm{R}}{\mathrm{kpc}}\right) .
\end{equation}
Using Equation \ref{eqn: m_dyn general} with the half light radius of Sérsic 1, we estimate a dynamical mass of $\mathrm{log}_{10} (M_\mathrm{dyn}/M_\odot) = 9.35\pm0.43$, giving a stellar mass fraction of $\sim18\%$, consistent with predictions from simulations \citep{de_Graaff_TNG_24}.

\section{Discussion} 
\label{sec: discussion}
\subsection{Physical Properties}
\begin{figure}
    \centering
    \includegraphics[width=0.9\columnwidth]{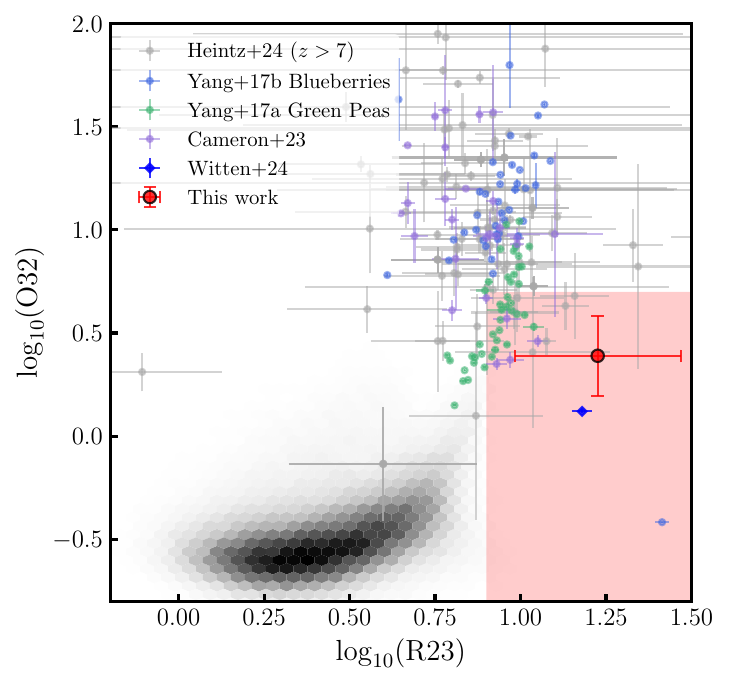}
    \caption{Dust corrected O$_{32}$-R$_{23}$ plot, showing GNWY-7379420231 compared to NIRSpec data \citep{Heintz2024, cameron_jades_2023, Witten24}, local analogues \citep{Yang_green_pea, Yang_blueberries}, and the Sloan Digital Sky Survey \citep[SDSS;][]{SDSS} Data Release 7 \citep{sdss_dr7}, shown in black. The red shading shows the region $\log_{10}\mathrm{O}{32} < 0.7$ and $\log_{10}\mathrm{R}{23} >0.9$ which may contain galaxies hosting an older stellar population \citep{Witten24}.}
    \label{fig:O32_R23}
\end{figure}

The physical properties of \id\ can provide insights into the formation and evolution of our galaxy, allowing us to discuss this source in the wider context of galaxy and dust formation in the early universe. 
From the integrated PRISM spectrum of our source, we measure a high [O\textsc{iii}]+H$\beta$ rest-frame equivalent width of $1110\pm183$\AA, placing GNWY-7379420231 within the extreme emission line galaxy (EELG) regime \citep[e.g.,][]{Boyett_2024}. 
The [O\textsc{iii}]+H$\beta$ emission is a tracer of ongoing star formation, indicating the presence of a young stellar population. This is in agreement with the inferred mass-weighted ages ($22-59$ Myr) from the combined spectro-photometric SED fitting, when a UV bump is included in the attenuation curve (see Table \ref{tab:bagpipes_comp}). 
We also measure the dust-corrected emission line ratios $\mathrm{O}{32}$ and $\mathrm{R}{23}$, which are shown in Figure \ref{fig:O32_R23}. Our source falls within the red shaded region defined in \citet{Witten24}, which indicates the region in the $\mathrm{log}_{10}(\mathrm{O}{32}) - \mathrm{log}_{10}(\mathrm{R}{23})$ parameter space that may be populated by galaxies containing an older stellar population. However, in galaxies with strong emission lines, the light from recent starbursts can dominate that of older stellar populations in an effect known as `outshining' \citep{Narayanan_outshining}. Due to the presence of extreme emission lines in the integrated spectrum, we investigate whether an older stellar population is present through both resolved SED fitting in Voronoi bins, and the SED fitting of the \textsc{galfit} components.

Through morphological analysis we are able to separate the source into three components: the main component (Sérsic 1) which contains the bulk of the stellar mass, and a tail made up of a second Sérsic component (Sérsic 2) and a point source. The inferred mass-weighted ages from the SED fitting of these components suggests that the main component is significantly older, with $t_\star \sim 252$ Myr, compared to $t_\star\sim56$ Myr and $t_\star\sim3$ Myr for the components within the tail. This indicates that an older stellar population is indeed present within GNWY-7379420231, despite the extreme emission lines dominating the integrated spectrum. The best-fit posterior spectra are shown in Figure \ref{fig:galfit_bagpipes}, with the UV bump feature strongest in the point source component of the tail. 

The three components identified in the morphological analysis may correspond to three distinct regions within {\id}, where Sérsic 1 corresponds to the older stellar population with the bulk of the stellar mass, the point source corresponds to the merging clump, and Sérsic 2 is the resulting tidal tail. In order to discuss the implications of the minor merger, we isolate the merging component by identifying the region of the southern Sérsic component (Sérsic 2) that does not show any overlap with the northern Sérsic component (Sérsic 1). The isolated region includes both the point source and tidal tail, and is henceforth referred to as the merging component. This process is outlined in Appendix \ref{morph_appendix} with the merging component shown in Figure \ref{fig:merger_component_map}.

We show maps of the physical properties inferred from the resolved SED fitting in Figure \ref{fig:resolved}, and UV continuum slopes measured from the median \textsc{bagpipes} posterior spectra. 
The overlaid contours from the SW stack show the clumpy nature of the galaxy, with a main component and an extended tail-like feature. Although clumpy morphologies are common within the EoR \citep{Chen2024}, the dissimilar star formation histories of each component suggest that this may be a merger system \citep[e.g.,][]{Hsiao2023}. Furthermore, this hypothesis is supported by the detection of a tail-like feature \citep[e.g.][]{Ren2020}. 

The regions with the highest [O\textsc{iii}]+H$\beta$ EWs ($>2500$\AA) are located across the merging component, indicating that the starburst is merger-induced. The mass-weighted ages inferred by the SED fitting for this region of extreme line emission are very young ($\lesssim 20$ Myr), further supporting the idea of a recent burst of star formation. 
The stellar mass surface density ($\Sigma_\star$) map shows that the bulk of the stellar mass is concentrated within the main component of the source. This could suggest the presence of an older stellar population in this region where stellar mass has built up over time. Although the $\Sigma_\star$ and star formation rate surface density ($\Sigma_{\mathrm{SFR}_{10}}$) overlap significantly, there is a slight offset between the well localized peaks of the $\Sigma_{\mathrm{SFR}_{10}}$ and $\Sigma_\star$, with the peak $\Sigma_{\mathrm{SFR}_{10}}$ slightly closer to the merging component. Through the maps of the UV slope and $A_V$, we can see that the dust attenuation is patchy, with some significantly dustier sightlines present. The location of these dustier sightlines suggests that there is increased dust build up localized to the merging component, which has $A_V \sim 0.26$ mag, compared to $A_V \sim 0.19$ mag in the main component. The bump strength ($B$) parameter from the Salim dust curve is a measure of the additional dust attenuation at $2175$\AA\ and peaks within the merging component, spatially aligned with the region of extreme line emission where a recent burst of star formation took place. Note that we aim to be conservative in our resolved bump fitting by adding a prior on $B$ centred on $B=0$ (see Sections \ref{sec:morph} and \ref{sec:resolved}).  While our photometric analysis provides an initial insight into this spatial distribution, we note that observations with the NIRSpec IFU would be valuable to explore these findings in greater detail.

There are two possible scenarios which may give rise to the visibility of the UV bump in the merging component of {\id}. 
Firstly, an older stellar population may have enriched this region over time. The total stellar mass formed $>300$ Myr ago within the main Sérsic component (Sérsic 1) is $\mathrm{log}_{10}(M_{\star}/M_\odot) = 7.96^{+0.12}_{-0.21}$. It is possible that AGB stars from this epoch could have contributed to the build up of carbonaceous dust grains, thus contributing to the presence of the UV bump feature. In this case, the merger-induced starburst would illuminate existing dust.
Alternatively, the merger-induced starburst could have processed early-formed dust, breaking down larger dust grains formed in SNe, into smaller carbonaceous particles responsible for the UV bump (see Section \ref{sec:dust production}).
Although some studies suggest that small dust grains are destroyed in hard radiation fields, which leads to flatter attenuation curves with weaker UV bumps \citep[e.g.,][]{kriek_conroy_2013, tazaki_2020}, recent \emph{JWST} observations have shown that PAHs can survive even in the harsh environments near active galactic nuclei (AGN) \citep{GATOS_24}. As such, it remains unclear whether dust destruction, or the processing of larger grains into smaller grains, is the dominant process in hard radiation fields.

We explore the ISM properties of GNWY-7379420231 with the high-resolution G395H spectrum, dominated by emission from the merging component of the galaxy, which we use to estimate the electron density. While the obtained value is high compared to estimates based on the O${32}$ emission line ratio \citep{Reddy_2023_electron_density}, and elevated compared to galaxies at lower redshift \citep[e.g. $n_e \sim 100 \mathrm{cm}^{-3}$,][]{kaasinen2017}, it is in agreement with the values obtained for galaxies at a similar redshift in \citet{Isobe2023}. The high electron density ($\log_{10}(n_e)$ cm$^{-1}$ = $3.12^{+0.83}_{-0.47}$) may be connected with gas compression as a result of the merger, potentially inducing the recent burst of star formation. Finally, we estimate the dynamical mass of our galaxy, finding a stellar mass fraction of around $\sim18\%$, suggesting a gas dominated system. The gas supplied by the merger would be required to fuel the strong burst of star formation.
However, it must be noted that dynamical masses may be under or overestimated in case of a merger \citep{Kohandel2019, de_graaff_lsf}.

Finally, {\id} is kinematically coincident with two overdensities (JADES-GN-OD-7.133, JADES-GN-OD-7.144) identified in \citet{Helton2024}, which reside in a complex environment with connected filamentary structures. Furthermore, there is evidence of accelerated galaxy evolution in protocluster environments \citep{Morishita_24}, suggesting that the large scale environment in which {\id} resides could influence its evolution, contributing to rapid dust formation.

\subsection{Dust Attenuation Curves}
\label{sec: dust curves}
The dust attenuation curve assumed during SED fitting can introduce systematic biases, with properties such as stellar mass, SFR, and $A_V$ varying significantly \citep{kriek_conroy_2013, Salim_2020}. Previous studies have shown that stellar masses and SFRs may vary by up to $0.16$ dex and $0.3$ dex, respectively \citep{Reddy_2015, Narayanan_2018, Tress_2018, Shivaei_reddy_2020}, indicating that care must be taken when carrying out SED fitting. In this section, we explore the impact of four different dust attenuation curves, three of which incorporate a UV bump.

We find that the inferred metallicity ($Z_\star$) and ionization parameter ($\mathrm{log}_{10}U$) are consistent regardless of the dust law assumed during SED fitting, in agreement with \citet{Markov23}. While the median mass-weighted age, $t_\star$, varies from $22 - 166$ Myr, the large uncertainties associated with these inferred values have large overlap, suggesting the ages are broadly consistent regardless of the assumed dust law. 
However, we find that the stellar mass, $\log _{10}\left(M_{\star} / M_{\odot}\right)$, does vary depending on the assumed dust law. Adopting the flat Salim dust law gives rise to a higher stellar mass ($\log _{10}\left(M_* / M_{\odot}\right) = 8.84^{+0.13}_{-0.15}$, compared to $\log _{10}\left(M_* / M_{\odot}\right) = 8.60^{+0.15}_{-0.13}$ when assuming the standard Salim dust law). This contrasts with the consistent values inferred when adopting a dust curve which exhibits a UV bump, due to the flat Salim dust law not accounting for the deficit in the rest-UV flux caused by the UV bump, thus causing the stellar mass to be overestimated. This is also seen in \citet{fisher2025rebelsifudustattenuationcurves} in sources with a strong UV bump. Additional effects of not accounting for the UV bump are documented in the literature, with changes in the estimated UV continuum slopes and dust-corrected SFRs varying by up to an order of magnitude \citep[e.g.][]{Buat2011, Narayanan_2018, Tress_2018, Shivaei_reddy_2020, Shivaei_2022}. Therefore, care should be taken when interpreting stellar population parameters derived assuming a bump-free attenuation curve in such systems. 

The SFRs and V-band dust attenuation also vary, with the use of the Li parameterization resulting in higher inferred values for both quantities.

The dust attenuation curves obtained from the \textsc{bagpipes} fitting are shown in Figure \ref{fig:dust}, along with the commonly used Calzetti, MW, and SMC curves. We find that the dust curve obtained using the \citet{Salim2018} curve is similar to the MW dust curve within $1\sigma$ errors,  with a strong UV bump at $2175$\AA, with a bump strength of $B=4.08^{+1.42}_{1.08}$ and a power-law modification of $\delta = -0.11^{+0.13}_{-0.16}$. The dust curve obtained with the Li parameterization most resembles the shape of the MW curve with the presence of the UV bump, compared to other curves such as the Calzetti and SMC curves. The dust curve obtained using the flat Salim dust curve closely resembles the Calzetti curve ($\delta = 0.03^{+0.09}_{-0.13}$). 

{\renewcommand{\arraystretch}{1.5}
\begin{table}
    \caption{Best-fit values for the physical properties of GNWY-7379420231 from SED fitting with differing dust attenuation curves. The first row contains the 10 Myr SFRs, the second row provides the mass weighted ages, the third row gives the stellar metallicity, the fourth row the stellar mass, the fifth row the ionization parameter, and the final row gives the $V$ band dust attenuation. All errors are the $16$th and $84$th percentiles of the posterior distribution.}
    \centering
    \resizebox{0.99\columnwidth}{!}{
    \begin{tabular}{l|c|c|c|c}
      Parameter   &  Salim & Li & MW & \begin{tabular}{@{}c@{}}Salim \\ ($B=0$)\end{tabular} \\ \hline
        SFR$_{10}$ (M$_\odot$ yr$^{-1}$) & $15.5^{+7.4}_{-5.2}$ & $24.7^{+7.0}_{-8.5}$ & $16.2^{+7.5}_{-4.8}$ & $11.3^{+4.5}_{-3.2}$\\
        $t_*$ (Myr) & $59^{+125}_{-42}$ & $22^{+61}_{-16}$ & $39^{+85}_{-24}$ & $166^{+105}_{-99}$\\
        $Z_\star (Z_\odot)$ & $0.27^{+0.06}_{-0.05}$ & $0.29^{+0.06}_{-0.06}$ & $0.32^{+0.03}_{-0.03}$ & $0.32^{+0.08}_{-0.10}$\\
        log$_{10}(M_\star/M_\odot)$ & $8.60^{+0.15}_{-0.13}$ & $8.57^{+0.13}_{-0.10}$ & $8.63^{+0.13}_{-0.12}$ & $8.84^{+0.13}_{-0.15}$\\
        log$_{10}U$ & $-1.73^{+0.15}_{-0.15}$ & $-1.76^{+0.17}_{-0.17}$ & $-1.77^{+0.16}_{-0.15}$ & $-1.83^{+0.15}_{-0.13}$\\ 
        $A_V$ (mag) & $0.27^{+0.06}_{-0.05}$ & $0.40^{+0.05}_{-0.06}$ & $0.32^{+0.03}_{-0.03}$ & $0.31^{+0.08}_{-0.10}$\\ 
        \hline
    \end{tabular}}
    \label{tab:bagpipes_comp}
\end{table}}

\subsection{The 2175\AA\ UV Bump}
\label{sec: bump discussion}

\begin{figure}
    \centering
    \includegraphics[width=0.99\columnwidth]{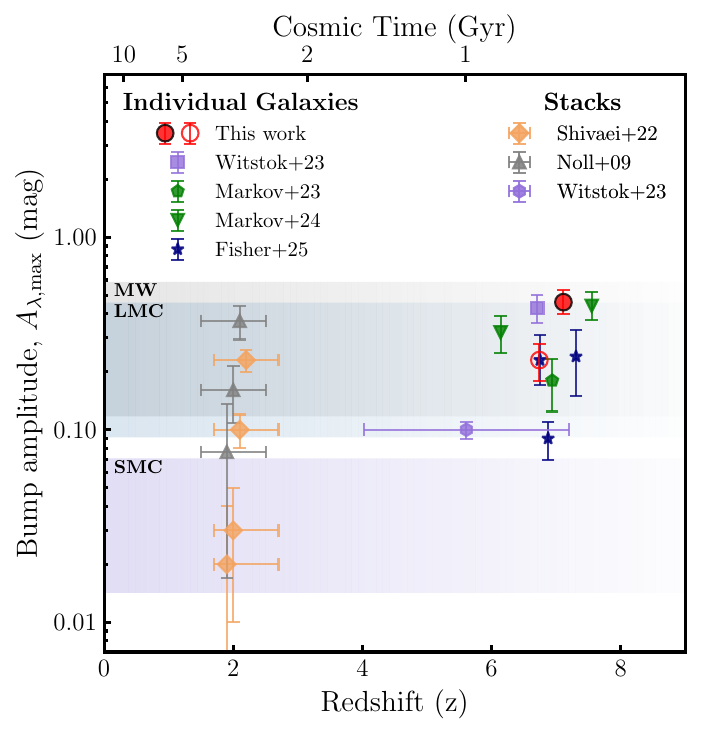}
    \caption{The bump amplitude, $A_{\lambda,\mathrm{max}}$ as a function of redshift for individual high redshift sources which show evidence for the presence of a UV bump \citep{Witstok_2023_Bump, Markov23, Markov24, fisher2025rebelsifudustattenuationcurves}, $z\sim2$ stacks \citep{Shivaei_2022, Noll_2009} and a stack of 10 $z\sim4-7$ galaxies which show evidence for a UV bump \citep{Witstok_2023_Bump}. The UV bump detected in  GNWY-7379420231 is shown by the solid red circle, and the tentative UV bump detected in EGSZ-9135048459 is shown by the open red circle. The points are staggered for clarity, and the error bars along the x-axis represent the full redshift range for each stack. The shaded regions represent the average bump amplitudes in the SMC, LMC, and MW extinction curves \citep{Fitzpatrick_1986, Gordon_2003} for $0.1 ~\mathrm{mag}~< A_V < 0.5 ~\mathrm{mag}$.}
    \label{fig:bump amplitude}
\end{figure}

Investigating the properties of the UV bump are crucial for understanding the evolution of cosmic dust grains. We measure a central wavelength of $\lambda_{\max } = 2257^{+26}_{-28}$\AA, similar to the $\lambda_{\max }=2263_{-24}^{+20}$\AA \ measured in \citet{Witstok_2023_Bump}. This is $\sim2.9\sigma$ higher than the peak wavelength seen in the MW curve, and may be caused by dust grains with larger molecular size \citep{blasberger2017, Li_2024_UV_Bump, Lin_2025}.

We measure a bump amplitude (strength) of $A_{\lambda,\mathrm{max}} = 0.46^{+0.06}_{-0.07}$ mag, which we plot against cosmic time in Figure \ref{fig:bump amplitude}, along with other high redshift detections of the UV bump \citep{Witstok_2023_Bump, Markov23, Markov24, fisher2025rebelsifudustattenuationcurves}, $z\sim2$ stacks \citep{Shivaei_2022, Noll_2009}, and the MW, LMC and SMC extinction curves \citep{Fitzpatrick_1986, Gordon_2003}. 
We note that many high-redshift galaxies do not show clear evidence for a UV bump \citep[e.g.][]{Markov23, Markov24, fisher2025rebelsifudustattenuationcurves}, and are therefore not included in this comparison. While the bump amplitudes in this work and \citet{Shivaei_2022, Witstok_2023_Bump} are measured in the same way, we must convert the other literature points to a consistent definition of $A_{\lambda,\mathrm{max}}$. We first convert the MW, LMC and SMC curves using the \citet{Fitzpatrick_1986} definition $A_{\lambda,\mathrm{max}} = c_3 /\gamma^2E(B-V)$, where we vary $E(B-V) = A_V/ R_V$ over a range $0.1~\mathrm{mag} < A_V < 0.5~\mathrm{mag}$. We convert the \citet{Noll_2009} values in the same way, with the measured values of $E(B-V)$.  We convert the \citet{Markov23} value by measuring the excess attenuation using their quoted values of $c_1$ to $c_4$ and the \citet{Li_2008} dust attenuation expression as defined in their Equation 6, compared to the baseline attenuation determined by setting $c_4 = 0$. 
Similarly, we convert the \citet{fisher2025rebelsifudustattenuationcurves} values by measuring the excess attenuation using the quoted values of $\delta$ and $B$ for the Salim dust law, and comparing to the baseline attenuation when $B=0$.
Finally, we download the spectra of the two sources identified in \citet{Markov24} with a bump detection from the DAWN \emph{JWST} Archive \citep[DJA;][]{Heintz_DLA}, which are reduced with \texttt{msaexp} \citep{msaexp, de_graaff_RUBIES} to measure the  bump strength and central wavelength, following the same procedure as in Section \ref{fig:UV fit}. We measure bump amplitudes for 2750\_449 and 1433\_3989 of $A_{\lambda,\mathrm{max}} = 0.44^{+0.07}_{-0.08}$ mag and $A_{\lambda,\mathrm{max}} = 0.32\pm0.07$ mag, respectively.

It is expected that the strength of the UV bump decreases towards higher redshift \citep{Markov24}, however the bump amplitude measured in this work is high and in contrast to this expected trend. Interestingly, it is similar to the $z \sim 6.7$ detection in \citet{Witstok_2023_Bump} and values we measure for the two galaxies from \citet{Markov24}. Combined with the increased peak wavelength of the bump feature in GNWY-7379420231, this suggests that the grain composition may differ compared to that at lower redshift, or these young galaxies could have a simpler dust-star geometry, as it is expected that the galaxies with the most complex young geometries have weaker bump strengths \citep{Narayanan_2018}.  

\subsection{Dust Production in the Early Universe}
\label{sec:dust production}
The detection of the $2175${\AA} UV bump in GNWY-7379420231 provides important constraints on its dust properties and evolution. The UV bump is predominantly seen in metal-rich galaxies at $z\lesssim3$ \citep[e.g.,][]{eliasdottir2009, Noll_2009, Shivaei_2022}, suggesting it is commonly found in evolved systems. However, recent studies have found no significant trend between gas-phase metallicity and the $2175${\AA} feature in local galaxies, across the metallicity range $8.40 < 12 + \log(\mathrm{O/H}) < 8.65$ \citep{Battisti_2025}.

We find that our source is metal-enriched compared to galaxies of a similar mass \citep{curti_jades_2024}, with $Z _\mathrm{neb}\sim 0.30Z_\odot$. The dust evolution within galaxies depends strongly on the age and metallicity of the system, with dust production in low metallicity systems controlled by stellar sources (AGB stars and SNe II). When the metallicity exceeds a critical metallicity ($Z_\mathrm{cr}$), dust mass growth becomes dominated by metal accretion onto existing dust grains within the ISM, and dust mass increases rapidly. This transition may occur at $10-20\%$ solar metallicity \citep[e.g.][]{Asano_2013, Remy_ruyer_2014, remy_ruyer_2015,Li_2019, Roman_Duval_2022}. The metallicity of our system suggests that it has entered the regime of efficient ISM dust mass build up. Additionally, \citet{Nanni_2025} recently explored dust production pathways for the UV bump galaxy identified in \citet{Witstok_2023_Bump}, finding that when dust growth in the ISM is included, $\sim85\%$ of the carbonaceous dust production is due to dust growth. 

From our SED fitting analysis of the integrated spectrum when adopting the Li dust model, we infer the presence of a very young stellar population, with $t_\star \sim 22$ Myr. If we were to rely solely on the stellar age inferred from the integrated spectrum in isolation, we must consider alternative dust production pathways, given that AGB stars capable of producing carbonaceous dust require a $\sim300$ Myr timescale to evolve off the main sequence. 

A potential mechanism is through Wolf-Rayet (WR) stars, formed when massive stars with initial masses $> 30~M_\odot$ lose their hydrogen envelope. Carbon sequence WR stars (WC stars) are known to produce dust, including PAHs \citep{Lau_2022}, although they may need to be in a binary system where the companion has a high mass-loss rate \citep[e.g.,][]{cherchneff_2000, Lau_2021,peatt_2023, schneider_formation_2023}. However, their contribution may be limited: just $27\pm9\%$ of WC stars display circumstellar dust within the Milky Way \citep{Rosslowe_2015}, and WR stars are rare \citep{Eldridge_2017}, with few WC stars found in low metallicity environments \citep{Massey_2003}. Nonetheless, the large $0.1-1.0\mu$m grains produced would be more robust to destruction from the subsequent SN shocks, and have grain lifetimes $\sim3$ times greater than $100$\AA\ sized grains \citep{Jones_1996}. PAHs are also known to be highly stable due to their honeycomb structure \citep{allamandola_1989, Tielens_2008, Lau_2022}.

Alternatively, early dust production could be dominated by SNe unless the reverse shock is very significant, even more so if the IMF is top-heavy \citep[][and references therein]{schneider_formation_2023}. Type II supernovae produce dust primarily made up of silicates, amorphous carbon, magnetite, and corundum \citep{todini_2001}, which can be processed into PAHs. Amorphous carbon ejected into the ISM can react with hydrogen to form hydrogenated amorphous carbons (HACs), which can then form PAHs through shattering due to grain-grain collisions \citep[][]{Jones_1996}. Additionally, photoprocessing by UV radiation can lead to the formation of aromatic bonds, with larger carbonaceous particles acting as a reservoir for the formation of smaller particles \citep{Duley_2015}. Furthermore, graphitic grains with isotopic compositions that suggest an origin in SNe have been identified, most consistent with an origin in Type II SNe \citep[e.g.,][]{Zinner_1998, Nittler_2016}. While the SNe reverse shock may preferentially destroy smaller grains \citep{Nozawa_2007}, \citet{Jones_1996} suggests that as much as $5-15\%$ of the starting graphite grain mass may end up in $<14$\AA\ graphitic fragments, potentially forming PAHs through hydrogenation. 

However, morphological analysis of our source reveals a more complex system than initially suggested by the integrated spectrum. When we examine the source as a three-component system, we find evidence for an older stellar population masked by outshining effects, a particular problem when coverage is limited to the rest-frame UV and optical \citep{Giminez_Arteaga_outshining}. This older stellar population is concentrated within Sérsic 1, the most massive of the three components, with a mass-weighted age of $t_\star = 252^{+20}_{-46}$ Myr. 
While stars across a broader mass range ($0.8$-$8M_\odot$) can enter the AGB, carbon grains are mostly produced in AGB stars within the mass range $2M_\odot < m_\mathrm{star} < 3-3.5M_\odot$ \citep{schneider_formation_2023}.
Crucially, this component shows substantial stellar mass build up at ages $> 300$ Myr, where stars $\sim3M_\odot$ enter the AGB. This could provide sufficient time for AGB driven dust production to pre-enrich the ISM before the merger event, although their evolution is less well understood in low-metallicity environments \citep[e.g.,][]{Hewig_2005}.
While the SED fitting of the \textsc{galfit} components shows that the UV bump is weakest is the oldest component, this does not prevent it from playing an important role in the presence of the UV bump. Its age and stellar mass suggest a scenario in which it may have pre-enriched the surrounding ISM with carbonaceous dust grains. The contribution of this early dust build up may now be observable, with the merger-induced starburst illuminating, or reprocessing (larger) pre-existing carbonaceous dust grains and allowing for the detection of the UV bump.

Recent galaxy evolution simulations \citep{Naryanan_2023_PAH, Narayanan_2024_UV_Slopes} suggest that PAH formation is enhanced in environments with high velocity dispersions (highly turbulent gas) and strong radiation fields.
Increased shattering rates are driven by these large ISM velocity dispersions in galaxies with high sSFRs, resulting in increased feedback energy per unit mass, driving up the fraction of ultrasmall grains. Furthermore, elevated global SFRs can drive aromatization by UV radiation \citep{Naryanan_2023_PAH}. These theoretical predictions align with our observations of GNWY-7379420231, where the UV bump is localized to the merging component characterized by extreme [O\textsc{iii}]+H$\beta$ equivalent widths and sSFRs. This spatial alignment, combined with the evidence for an older stellar population in the main component, suggests that the intense UV radiation and turbulence in the merging component may be driving localized PAH formation through both top-down shattering, as well as increased dust growth on small grains within the turbulent ISM \citep{Narayanan_2024_UV_Slopes}.

\section{Summary}
\label{sec:summary}
In this paper, we have presented the analysis of \emph{JWST}/NIRSpec observations, revealing one of the most distant known galaxies exhibiting a very strong $2175$\AA\ UV bump feature at $z=7.11$ when the Universe was only $\sim 700$ Myr old. We have presented a detailed analysis of the dust properties and stellar populations within GNWY-7379420231, finding evidence for both intense ongoing star formation and an older stellar population, suggesting a complex dust production history. The spatial correlation between the UV bump and the recent burst of star formation, combined with the system likely being a merger, provides new insights into high redshift dust evolution. Our main findings are summarized as follows:

\begin{itemize}
    \item We find a strong UV bump with $A_{\lambda,~\mathrm{max}} = 0.46^{+0.06}_{-0.07}$ mag in a galaxy at $z=7.11$. The peak wavelength of the UV bump is shifted by $84${\AA} (at $2.9\sigma$ significance) compared to that of the bump seen in the MW curve, which may suggest a different dust grain size distribution at high redshift.
    \item While the integrated spectrum suggests a young stellar population with $t_\star\sim22$ Myr, morphological analysis reveals the presence of an older stellar population in the most massive component with significant stellar mass with age $> 300$ Myr, potentially resolving the apparent tension in dust production timescales. 
    \item Through resolved SED fitting, we determine that the UV bump is spatially correlated with the merging component, characterized by extreme [O\textsc{iii}]+H$\beta$ equivalent widths, suggesting enhanced PAH formation in a turbulent environment with intense UV radiation. 
    \item We find that {\id} is metal enriched compared to galaxies of a similar mass, which could indicate rapid dust mass build up through dust grain growth mechanisms. The presence of metal enrichment could indicate that grain growth has reached an efficient growth regime, with the metal enrichment resulting from the underlying older stellar population.

\end{itemize}

In summary, we are able to provide new insights into potential dust formation and processing pathways at high redshift, suggesting that the ISM was pre-enriched by the older stellar population, before the dust was processed into PAHs through turbulence and UV radiation within the merging component. 
The capabilities of \emph{JWST} will allow us to probe dust production in the early universe in more depth, further constraining the early methods of dust production. Finding more galaxies which exhibit a UV bump, and building up a sample within the EoR will allow us to gain an understanding of the production mechanisms, the dust composition, and enable us to probe star formation and chemical evolution within the first billion years of cosmic time. Furthermore, upcoming sub-mm follow up will provide us with multi-wavelength dust constraints, such as providing an estimate of the dust-to-gas ratio of this source.

\section*{Acknowledgements}

The authors would like to thank the anonymous referee for their comments, which have improved this manuscript.
The authors would like to thank Adam Carnall and Thomas Harvey for helpful conversations. 
This work is based on observations made with the NASA/ESA/CSA James Webb Space Telescope (JWST). The data were obtained from the Mikulski Archive for Space Telescopes at the Space Telescope Science Institute, which is operated by the Association of Universities for Research in Astronomy, Inc., under NASA contract NAS 5-03127 for JWST. These observations are associated with programmes 1181 and 1211. This study made use of Prospero high-performance computing facility at Liverpool John Moores University. 
This work made use of Astropy:\footnote{http://www.astropy.org} a community-developed core Python package and an ecosystem of tools and resources for astronomy \citep{astropy:2013, astropy:2018, astropy:2022}. Some of the data products presented herein were retrieved from the Dawn JWST Archive (DJA). DJA is an initiative of the Cosmic Dawn Center (DAWN), which is funded by the Danish National Research Foundation under grant DNRF140.
KO would like to thank the Science and Technology Facilities Council (STFC) and Faculty of Engineering and Technology (FET) at Liverpool John Moores University (LJMU) for their studentship.
JW gratefully acknowledges support from the Cosmic Dawn Center through the DAWN Fellowship. The Cosmic Dawn Center (DAWN) is funded by the Danish National Research Foundation under grant No. 140. RS acknowledges support from a STFC Ernest Rutherford Fellowship (ST/S004831/1). 
MVM is supported by the National Science Foundation
via AAG grant 2205519. AJB, JC, acknowledge funding from the "FirstGalaxies" Advanced Grant from the European Research Council (ERC) under the European Union’s Horizon 2020 research and innovation programme (Grant agreement No. 789056) S.C acknowledges support by European Union’s HE ERC Starting Grant No. 101040227 - WINGS. BER acknowledges support from the NIRCam Science Team contract to the University of Arizona, NAS5-02015, and JWST Program 3215. 
RM acknowledges support by the Science and Technology Facilities Council (STFC), by the ERC through Advanced Grant 695671 “QUENCH”, and by the UKRI Frontier Research grant RISEandFALL. RM also acknowledges funding from a research professorship from the Royal Society.
ST acknowledges support by the Royal Society Research Grant G125142.

\section*{Data Availability}

This work is based on observations made with the NASA/ESA/CSA \emph{JWST}. The data were obtained from the Mikulski Archive for Space Telescopes at the Space Telescope Science Institute, which is operated by the Association of Universities for Research in Astronomy, Inc., under NASA contract NAS 5-03127 for JWST. These observations are associated with programmes PID 1211 (PI: K. Isaak) and PID 1181 (PI: D. Eisenstein).

\bibliographystyle{mnras}
\bibliography{paper} 



\appendix
\section{A tentative UV Bump detection in EGSZ-9135048459}
\label{egs appendix}
\begin{figure*}
    \centering
    \includegraphics[width=0.95\textwidth]{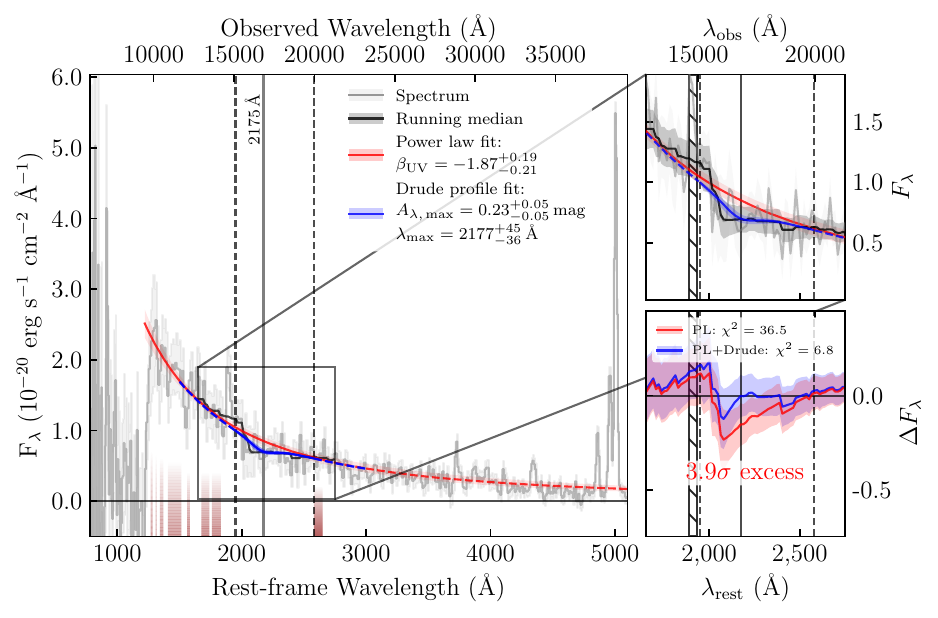}
    \caption{Spectrum of EGSZ-9135048459 (grey solid line) with a power-law fit to the UV continuum (red solid line). The dark red shading represents the UV
slope fitting windows. The zoom in panel of the region around 2175Å shows the running median, indicated by a solid black line. This represents the attenuated
stellar continuum, and shows a localized absorption feature. The Drude profile fit is shown by the solid blue line, within the fitting window indicated by the
vertical dashed lines. The hatched region shows the location of the  C\textsc{iii} doublet. The bottom right panel shows the residuals of the power-law fit (PL) and
the combined power-law and Drude profile fit (PL+Drude). The power-law fit alone has a $3.9\sigma$ negative flux excess.}
    \label{fig:egs_bump}
\end{figure*}

We detect a tentative UV bump in EGSZ-9135048459, at $z=6.74$, shown in Figure \ref{fig:egs_bump}. Following the methodologies detailed in Sections \ref{sec:UV mag slope} and \ref{sec:bump fitting}, we fit both the UV slope and UV bump feature. We do not include EGSZ-9135048459 in the main analysis due to the tentative nature of the detection, with a $3.9\sigma$ negative flux excess. We find that the measured peak wavelength, $\lambda_\mathrm{max}$, is consistent with that of the MW.

\begin{figure}
    \centering
    \includegraphics[width=0.99\columnwidth]{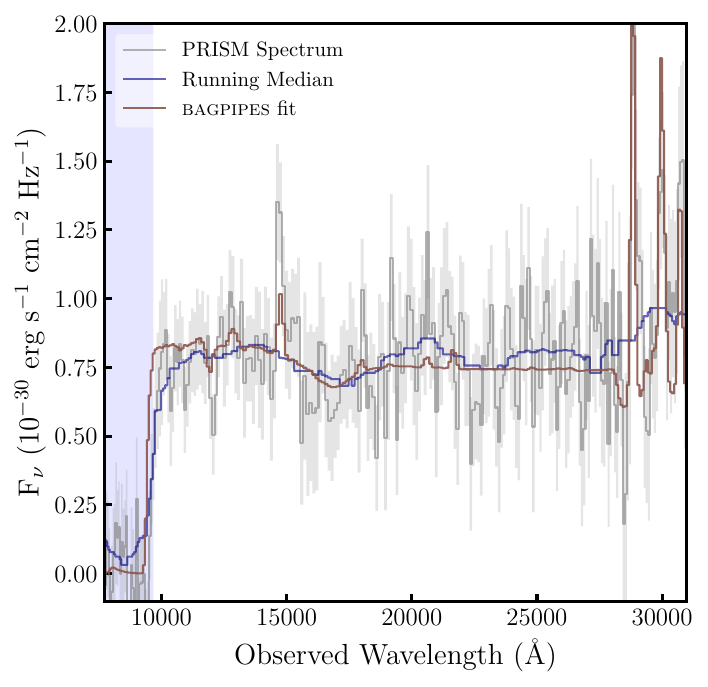}
    \caption{Posterior spectrum from \textsc{bagpipes} fitting of NIRSpec PRISM observations of EGSZ-9135048459. The observed spectrum and uncertainties are shown in grey, with the running median overlaid in dark blue. The best-fit \textsc{bagpipes} model spectrum is displayed in dark red.} 
    \label{fig:egs_bagpipes}
\end{figure}

We perform SED fitting of the NIRSpec PRISM spectroscopy only, following the methodology detailed in Section \ref{sec:sed fitting}. The best-fit \textsc{bagpipes} spectrum is shown in Figure \ref{fig:egs_bagpipes}, and shows tentative evidence for the presence of a UV bump. 

\section{SED Fitting}

The priors used for our \textsc{bagpipes} SED fitting in Section \ref{sec:sed fitting} are summarized in Table \ref{tab:priors}.

\begin{table}
    \centering
    \caption{Summary of SED fitting parameters and priors used in our \textsc{bagpipes} fitting in Section \ref{sec:sed fitting}. The minimum and maximum values allowed are given in brackets. The top section gives the shared parameters, which are the same regardless of the dust attenuation curve used. The following sections give model-specific parameters used with certain dust attenuation laws.}
    \begin{tabular}{l|l}
         Parameter & Prior  \\ \hline
         \multicolumn{2}{l}{Shared Parameters} \\ \hline
         $z$ & Fixed to 7.11235  \\ 
         $\log_{10}(M_\star/M_\odot)$ & uniform: (5, 12) \\
         $Z_\star/Z_\odot$ & uniform: (0, 0.5)  \\
         $\log_{10}U$ & uniform: (-3, -0.5)  \\ 
         $A_V$ (mag) & Gaussian: (0, 7), $\mu = 0.15, \sigma=0.15$ \\
         $\sigma_v$ (km s$^{-1}$) & logarithmic: (1, 1000) \\ 
         $\Delta\log$(SFR)$_i$ & Student's-t: (-50, 50) \\ \hline
         \multicolumn{2}{l}{Salim dust law} \\ \hline
         $B$ & uniform: (0, 10) \\ 
         $\delta$ & uniform: (-0.5, 0.2) \\ \hline 
         \multicolumn{2}{l}{Flat Salim dust law} \\ \hline
         $B$ & Fixed to 0 \\ 
         $\delta$ & uniform: (-0.5, 0.2) \\ \hline
         \multicolumn{2}{l}{Li dust law} \\ \hline
         $c_1$ & uniform: (0, 50) \\
         $c_2$ & uniform: (0, 20) \\
         $c_3$ & uniform: (-5, 75) \\
         $c_4$ & uniform: (0, 10) \\ \hline
    \end{tabular}
    \label{tab:priors}
\end{table}

\begin{figure}
    \centering
    \includegraphics[width=0.99\columnwidth]{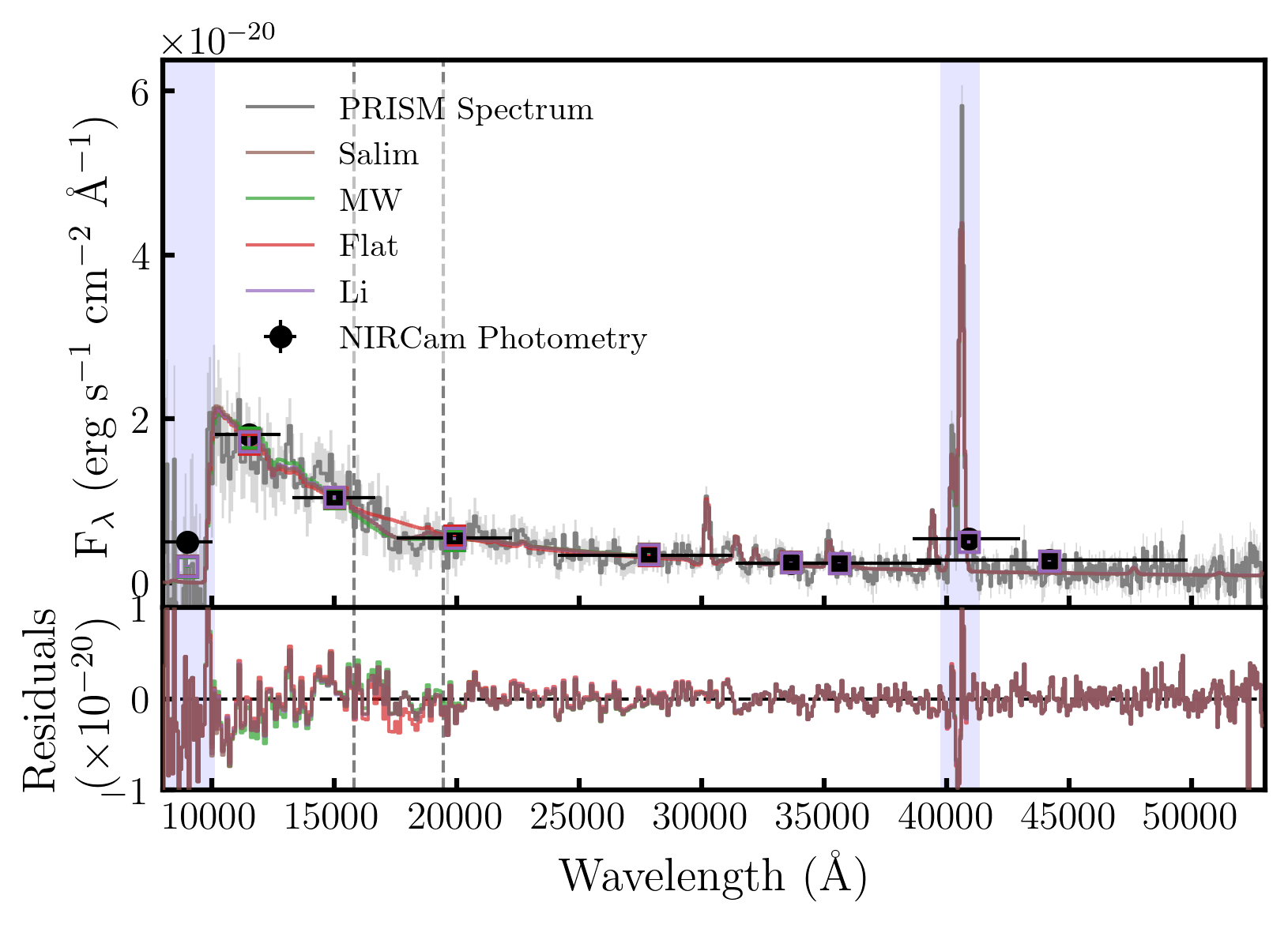}
    \caption{Top: Posterior spectra obtained from \textsc{bagpipes} fitting. The observed spectrum and associated errors are shown in grey, with the observed NIRCam photometry shown in black. The x error bars represent the filter width at $50\%$ of the maximum transmission. The posterior photometric points obtained from \textsc{bagpipes} are shown by open squares. Bottom: The residuals from the \textsc{bagpipes} fitting. The dashed vertical lines show the UV bump fitting region, and the blue shaded regions show the spectral regions masked in the \textsc{bagpipes} fitting.}
    \label{fig:bagpipes_models}
\end{figure}

We show the \textsc{bagpipes} posterior spectra obtained in Section \ref{sec:sed fitting}, with residuals, in Figure \ref{fig:bagpipes_models}. 

\begin{figure}
    \centering
    \includegraphics[width=0.99\columnwidth]{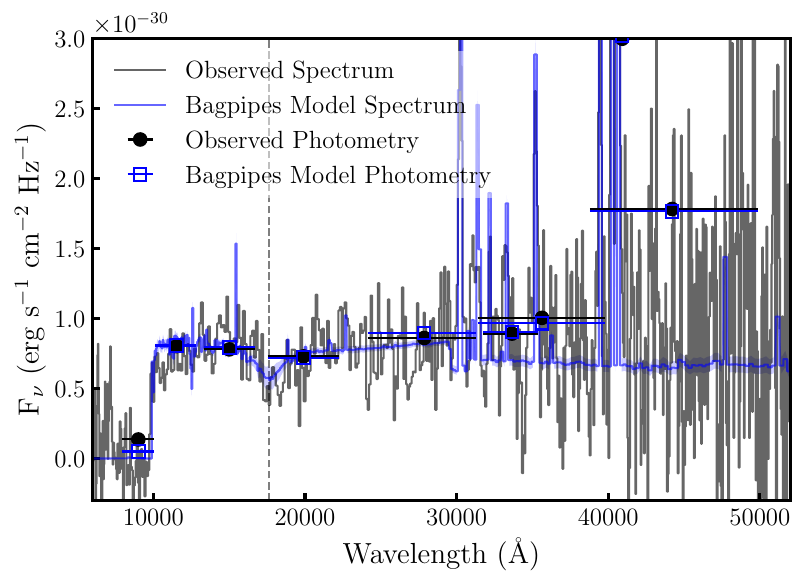}
    \caption{Posterior spectrum obtained from \textsc{bagpipes} fitting using the photometry only, shown in blue. The observed spectrum is shown in grey, with the observed NIRCam photometry shown in black. The x error bars represent the filter width at $50\%$ of the maximum transmission. The posterior photometric points obtained from \textsc{bagpipes} are shown by open blue squares. The dashed vertical line shows the location of the UV bump.}
    \label{fig:bagpipes_phot_only}
\end{figure}

In Figure \ref{fig:bagpipes_phot_only}, we show the \textsc{bagpipes} posterior spectrum obtained by fitting the photometry alone, following the method in Section \ref{sec:sed fitting}, with the Salim dust law. 

\section{Morphological fitting}
\label{morph_appendix}
\begin{figure*}
    \centering
    \includegraphics[width=1\linewidth]{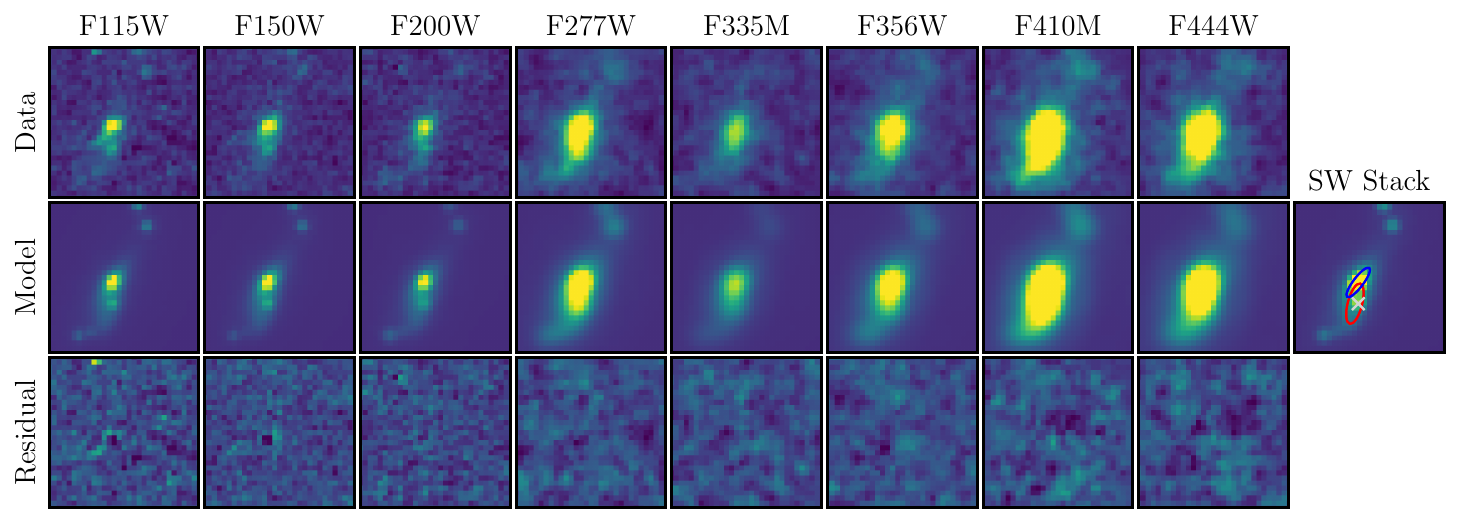}
    \caption{\textsc{galfit} fits to each band, excluding F090W. Top: the NIRCam cutout. Middle: the \textsc{galfit} model image. Bottom: the residual image (data-model). We show the best-fit model to the SW stack in the rightmost column for comparison, with the components overlaid. Sérsic 1 is shown in blue, Sérsic 2 in red, and the point source in the light grey cross. The data and model images in each band are linearly scaled between $-3-20\sigma$ of the data image background, and the residual images are scaled between $-3-10\sigma$ of the data image background, for clarity.}
    \label{fig:galfit_bands}
\end{figure*}

\begin{figure}
    \centering
    \includegraphics[width=0.99\columnwidth]{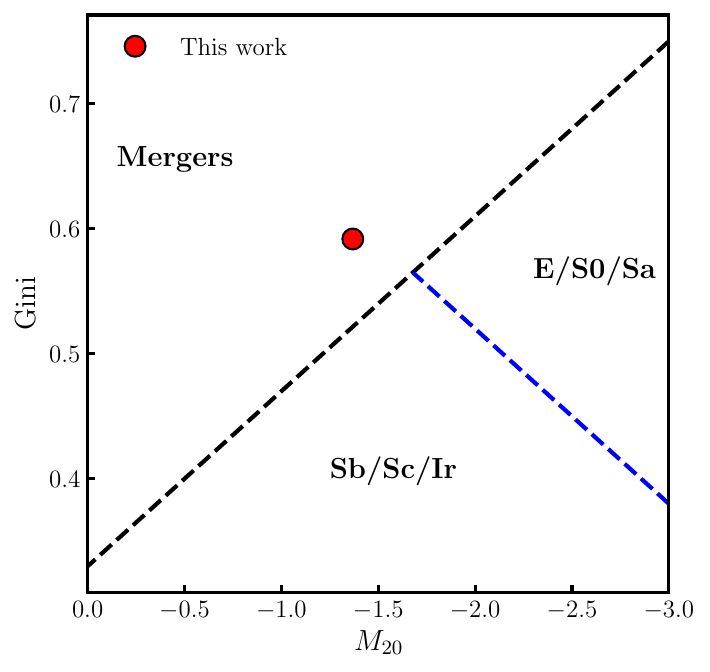}
    \caption{Gini vs $M_{20}$ for {\id}, measured using the SW stack. The black and blue dashed lines are from \citet{Lotz2008}, and show the division between merger candidates, E/So/Sa, and Sb/Sc/Ir galaxies.}
    \label{fig:gini_m20}
\end{figure}

\begin{figure}
    \centering
    \includegraphics[width=0.99\columnwidth]{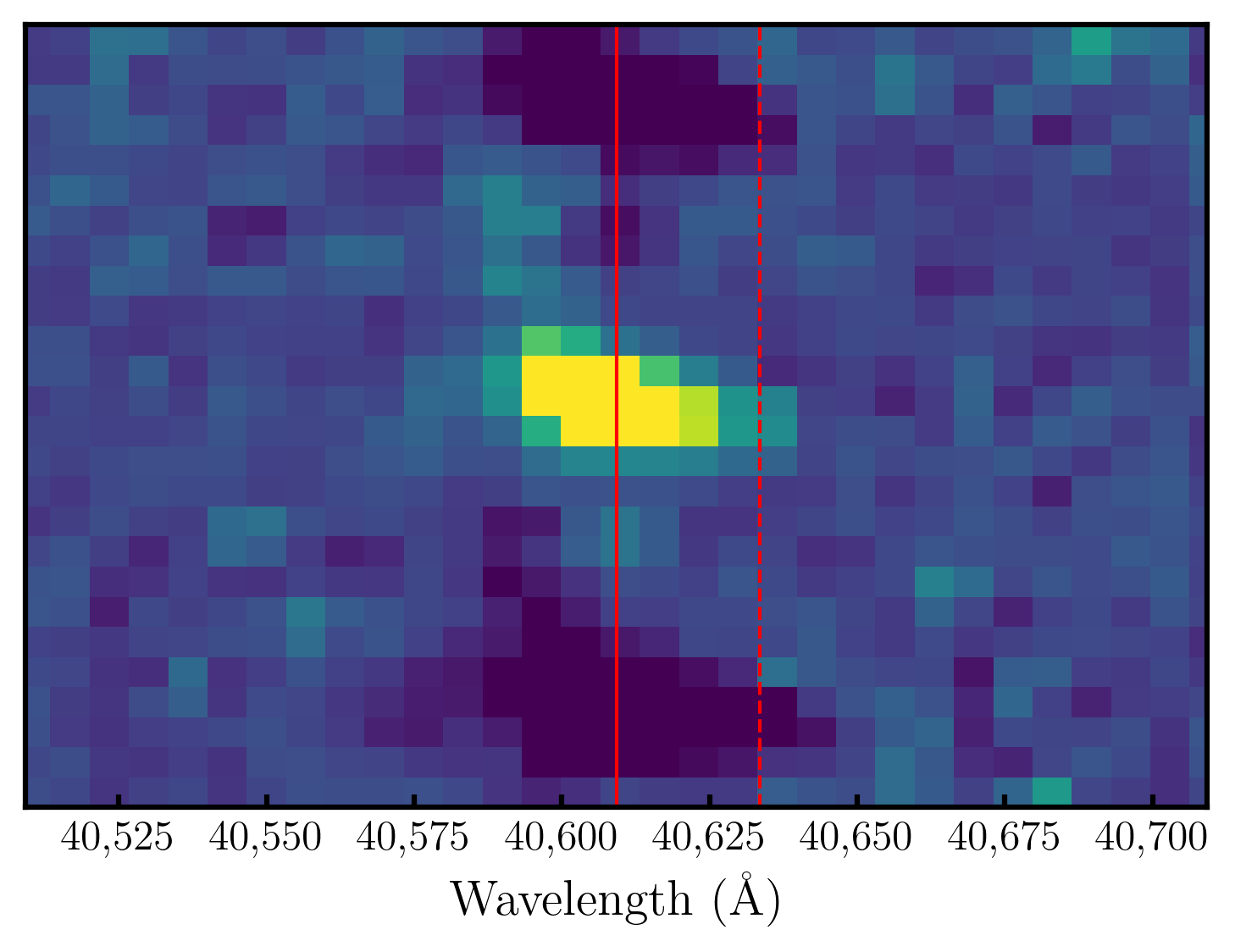}
    \caption{Zoom in on the [O\textsc{iii}] doublet in the G395H $R\sim2700$ grating. The solid red line shows the location of the primary [O\textsc{iii}]$\lambda5007$ emission, and the red dashed line indicates the offset component, likely originating from the merging tail of the galaxy.}
    \label{fig:vel offset}
\end{figure}

We fit each band with the best fit \textsc{galfit} model obtained in Section \ref{sec:morph}, leaving the magnitude free to vary. The data, model, and residual for each band is shown in Figure \ref{fig:galfit_bands}. We also show the location of {\id} in the $G-M_{20}$ parameter space in Figure \ref{fig:gini_m20}.  We show the G395H 2D spectrum of the [O\textsc{iii}] doublet in Figure \ref{fig:vel offset}.

\begin{figure}
    \centering
    \includegraphics[width=0.99\columnwidth]{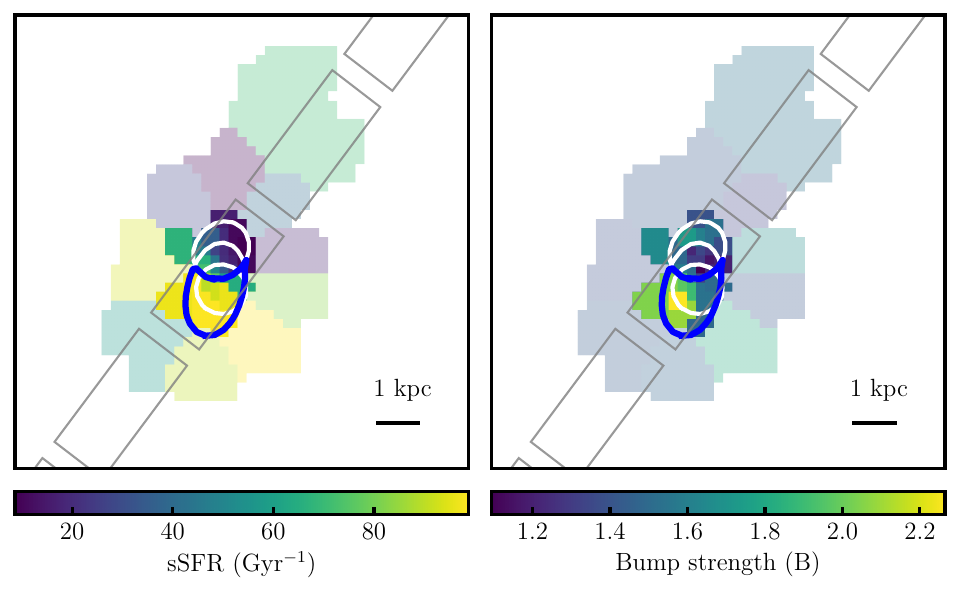}
    \caption{Maps of {\id}, with sSFR shown in the left panel, and bump strength ($B$) from the Salim dust law shown in the right panel. As in Figure \ref{fig:resolved}, the fainter bins are those where the photometry has at least one band with SNR $< 5$ (excluding F090W). The borders of each component are indicated by white lines, and the border of the merging component is indicated by the blue line. The NIRSpec slitlets are overlaid in grey. }
    \label{fig:merger_component_map}
\end{figure}

To define the region of the merging component, we first PSF match the best fit \textsc{galfit} model to the F444W image. We then define the boundary of each \textsc{galfit} component as the contour where the flux falls to $50\%$ of the component's maximum flux. The merging component is defined as the region of the southern Sérsic component (Sérsic 2) that does not overlap with the northern Sérsic component (Sérsic 1). The boundaries of each component and the merging component are shown in Figure \ref{fig:merger_component_map}.

\section{Emission Line Fluxes}
\label{electron_density_appendix}
\begin{figure}
    \centering
    \includegraphics[width=0.99\columnwidth]{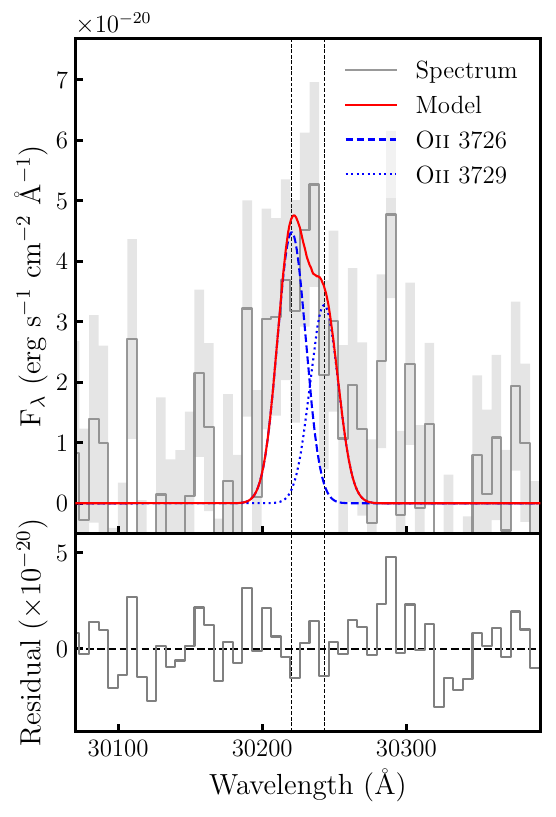}
    \caption{Double Gaussian fit to the O\textsc{ii} high-resolution G395H spectrum (grey solid line). The red solid line represents the overall model, convolved with the G395H LSF. The blue dashed and blue dotted lines represent the intrinsic fits to the individual emission lines. The bottom panel shows the residual between the spectrum and overall model. The black dashed lines show the wavelengths of the individual emission lines, which are fixed during the fitting process.}
    \label{fig:electron_density}
\end{figure}

Dust corrected emission line fluxes measured in Section \ref{sec:emission lines} are given in Table \ref{tab:line fluxes}. The [O\textsc{ii}] G395H emission line fit is shown in Figure \ref{fig:electron_density}.
\begin{table}
    \caption{Dust corrected emission line fluxes measured from the PRISM and G395H spectra. Fluxes are given in units of $10^{-18}$erg s$^{-1}$ cm$^{-2}$. [O\,{\sc ii}]$\lambda \lambda 3727\mathrm{,}29$ is blended in the PRISM spectrum, and H$\beta$ is located within the chip-gap in the G395H spectrum.}
    \centering
    \begin{tabular}{lcc}
         Emission Line & PRISM & G395H \\ \hline
         [O\,{\sc ii}]$\lambda \lambda 3727\mathrm{,}29$& $2.88 \pm 0.53$& -  \\
         $[\mathrm{O}\,\textsc{ii}]\lambda 3727$ & - & $1.10 \pm 0.28$ \\
         $[\mathrm{O}\,\textsc{ii}]\lambda 3729$ & - & $0.80 \pm 0.26$ \\
         H$\beta$ & $0.89 \pm 0.33$ & - \\
         $[\mathrm{O}\,\textsc{iii}]\lambda 4959$ & $2.94 \pm 0.11$ & $3.61 \pm 0.21$ \\
         $[\mathrm{O}\,\textsc{iii}]\lambda 5007$ & $8.76 \pm 0.32$ & $9.96 \pm 0.05$ \\ \hline
    \end{tabular}
    \label{tab:line fluxes}
\end{table}

\bsp	
\label{lastpage}
\end{document}